\documentclass[longauth]{aa}

\usepackage{comment}

\usepackage{natbib}
\bibpunct{(}{)}{;}{a}{}{,} 
\usepackage{amssymb}
\usepackage{url}
\usepackage{hyperref}
\usepackage{graphicx}
\usepackage{longtable}
\usepackage[usenames,dvipsnames]{color}
\usepackage{booktabs}
\usepackage{lscape}

\newcommand{\mm}{millimetre}

\newcommand{\msun}{\mbox{$M_\odot$}}
\newcommand{\mstar}{\mbox{$M_*$}}

\newcommand{\msunyr}{\mbox{\msun\ yr$^{-1}$}}
\newcommand{\hi}{H{\sc i}}
\newcommand{\mhi}{\mbox{$M_{\rm H{\sc I}}$}}
\newcommand{\zhi}{\mbox{$z_{\rm HI}$}}
\newcommand{\lphi}{\mbox{$L'_{\rm HI}$}}

\newcommand{\kms}{\mbox{km\,s$^{-1}$}}

\newcommand{\atcow}{AT\,2018cow}
\newcommand{\host}{CGCG\,137-068}
\urldef{\hyperleda}\url{http://leda.univ-lyon1.fr/ledacat.cgi?CGCG%20137-068}

\begin{document}

 \title{%
On the nature of the unusual transient {\atcow} from {\hi}~observations of its host galaxy
 }
 
\titlerunning{
Gas around {\atcow}
}
\authorrunning{Micha{\l}owski et al.}

\author{Micha{\l}~J.~Micha{\l}owski\inst{\ref{inst:poz}
}
\and
P.~Kamphuis\inst{\ref{inst:rub}}
\and
J.~Hjorth\inst{\ref{inst:dark}} 
\and
D.~A.~Kann\inst{\ref{inst:ant}}
\and
A.~de Ugarte Postigo\inst{\ref{inst:ant},\ref{inst:dark}}       
\and
L. Galbany\inst{\ref{inst:pitt}}
\and
J.~P.~U.~Fynbo\inst{\ref{inst:dawn}}     
\and
A.~Ghosh\inst{\ref{inst:aries}}
\and
L.~K.~Hunt\inst{\ref{inst:hunt}}                                                
\and
H.~Kuncarayakti\inst{\ref{inst:finca},\ref{inst:tuorla}}
\and
E.~Le Floc'h\inst{\ref{inst:sacley}}
\and
A.~Le\'sniewska\inst{\ref{inst:poz}}
\and
K.~Misra\inst{\ref{inst:aries},\ref{inst:davis}}
\and
A.~Nicuesa Guelbenzu\inst{\ref{inst:taut}}                      
\and
E.~Palazzi\inst{\ref{inst:pal}}                                         
\and
J.~Rasmussen\inst{\ref{inst:dark},\ref{inst:dtu}}      
 \and
L.~Resmi\inst{\ref{inst:iist}}            
\and
A.~Rossi\inst{\ref{inst:pal}}                                           
\and 
S.~Savaglio\inst{\ref{inst:sav}}                                                
\and
P.~Schady\inst{\ref{inst:bath}}    
\and
S.~Schulze\inst{\ref{inst:weiz}}
\and
C.~C.~Th\"one\inst{\ref{inst:ant}}
\and
D.~Watson\inst{\ref{inst:dawn}}
\and
G.~I.~G.~J\'ozsa\inst{\ref{inst:sarao},\ref{inst:dperu},\ref{inst:aia}}
\and
P.~Serra\inst{\ref{inst:oac}}
\and
O.~M.~Smirnov\inst{\ref{inst:dperu},\ref{inst:sarao}}
        }

\institute{
Astronomical Observatory Institute, Faculty of Physics, Adam Mickiewicz University, ul.~S{\l}oneczna 36, 60-286 Pozna{\'n}, Poland, {\tt mj.michalowski@gmail.com}\label{inst:poz}
\and
Astronomisches Institut der Ruhr-Universit\"{a}t Bochum (AIRUB), Universit\"{a}tsstrasse 150, 44801 Bochum, Germany\label{inst:rub}
\and
DARK, Niels Bohr Institute, University of Copenhagen, Lyngbyvej 2, DK-2100 Copenhagen \O, Denmark\label{inst:dark}
\and
Instituto de Astrof\' isica de Andaluc\' ia (IAA-CSIC), Glorieta de la Astronom\' ia, s/n, E-18008, Granada, Spain \label{inst:ant}
\and
PITT PACC, Department of Physics and Astronomy, University of Pittsburgh, Pittsburgh, PA 15260, USA \label{inst:pitt}
\and
The Cosmic Dawn Center, Niels Bohr Institute, University of Copenhagen, Lyngbyvej 2, DK-2100 Copenhagen \O, Denmark 
\label{inst:dawn}
\and
ARIES, Manora Peak, Nainital 263002 India \label{inst:aries}
\and
INAF-Osservatorio Astrofisico di Arcetri, Largo E. Fermi 5, I-50125 Firenze, Italy \label{inst:hunt}
\and
Finnish Centre for Astronomy with ESO (FINCA), University of Turku, V\"{a}is\"{a}l\"{a}ntie 20, 21500 Piikki\"{o}, Finland \label{inst:finca}
\and
Tuorla Observatory, Department of Physics and Astronomy, University of Turku, V\"{a}is\"{a}l\"{a}ntie 20, 21500 Piikki\"{o}, Finland \label{inst:tuorla}
\and
Laboratoire AIM-Paris-Saclay, CEA/DSM/Irfu - CNRS - Universit\'e Paris Diderot, CE-Saclay, pt courrier 131, F-91191 Gif-sur-Yvette, France \label{inst:sacley}
\and
Department of Physics, University of California, 1 Shields Ave, Davis, CA 95616-5270, USA \label{inst:davis}
\and
Th\"uringer Landessternwarte Tautenburg, Sternwarte 5, 07778 Tautenburg, Germany \label{inst:taut}\newpage
\and
INAF-OAS Bologna, Via Gobetti 93/3, I-40129 Bologna, Italy \label{inst:pal}
\and
Technical University of Denmark, Department of Physics, Fysikvej, building 309, DK-2800 Kgs. Lyngby, Denmark \label{inst:dtu}
\and
Indian Institute of Space Science \& Technology, Thiruvananthapuram 695547, India \label{inst:iist}
\and
Physics Department, University of Calabria, I-87036 Arcavacata di Rende, Italy \label{inst:sav}
\and
Department of Physics, University of Bath, Bath, BA2 7AY, United Kingdom \label{inst:bath}
\and
Department of Particle Physics and Astrophysics, Weizmann Institute of Science, Rehovot 7610001, Israel \label{inst:weiz}
\and
South African Radio Astronomy Observatory,
2 Fir Street, Black Rivetexr Park, Observatory, Cape Town, South Africa\label{inst:sarao}
\and
Department of Physics and Electronics, Rhodes University,
PO Box 94, Grahamstown, 6140, South Africa \label{inst:dperu}
\and
Argelander-Institut f\"ur Astronomie, Auf dem H\"ugel 71, 53121 Bonn, Germany \label{inst:aia}
\and
INAF - Osservatorio Astronomico di Cagliari, Via della Scienza 5, I-09047 Selargius (CA), Italy \label{inst:oac}
}

\abstract{%
Unusual stellar explosions represent an opportunity to learn about both stellar and galaxy evolution. Mapping the atomic gas in host galaxies of such transients can lead to an understanding of the conditions triggering them.
}
{We provide resolved atomic gas observations of the host galaxy, CGCG137-068, of the unusual, poorly-understood  transient {\atcow} searching for clues to understand its nature. We test whether it is consistent with a recent inflow of atomic gas from the intergalactic medium, as suggested for host galaxies of gamma-ray bursts (GRBs) and some supernovae (SNe).
}
{We observed the {\hi} hyperfine structure line of the {\atcow} host with the Giant Metrewave Radio Telescope.
}
{There is no unusual atomic gas concentration near the position of {\atcow}.
The gas distribution is much more regular than those of GRB/SN hosts. The {\atcow} host has an atomic gas mass lower by 0.24\,dex than predicted from its star formation rate (SFR) and is at the lower edge of the galaxy main sequence. 
In the continuum we detected the emission of {\atcow} and of a star-forming region in the north-eastern part of the bar (away from {\atcow}). This region hosts a third of the galaxy's  SFR.
}
{The absence of atomic gas concentration close to {\atcow}, along with a normal SFR and regular {\hi} velocity field, sets CGCG137-068 apart from GRB/SN hosts studied in {\hi}. 
The environment of {\atcow} therefore suggests that its progenitor may not have been
a massive star.
Our findings are consistent with an origin of the transient that does not require a connection between its progenitor and gas concentration or inflow: 
an exploding low-mass star, a tidal disruption event, 
a merger of white dwarfs, or a merger between a neutron star and a giant star. 
We interpret the recently reported atomic gas ring in CGCG\,137-068 as a result of internal processes connected with gravitational resonances caused by the bar.}

\keywords{dust, extinction -- galaxies: individual: CGCG\,137-068 --  galaxies: ISM -- galaxies: star formation -- supernovae: individual: AT 2018cow -- radio lines: galaxies}

\maketitle

\section{Introduction}
\label{sec:intro}

Unusual, luminous, and rare stellar explosions provide an opportunity to learn about stellar evolution and also about galaxy evolution in a broader context. An example of the latter approach is the possibility to select galaxies that experience a recent inflow of gas from the intergalactic medium (IGM) using host galaxies of long gamma-ray bursts (GRBs) and some types of supernovae (SN). Atomic gas concentrations away from the galaxy centres towards GRB/SN positions suggest an external origin of the gas \citep{michalowski15hi,michalowski16,michalowski18}, and a potential deficiency in molecular gas \citep{hatsukade14,stanway15,michalowski16,michalowski18co}. Studying gas inflows in such a direct way is important because they are required to fuel star formation in all galaxies, as implied from observations \citep{sancisi08,sanchezalmeida14b,spring17,elmegreen18b,combes18} and simulations \citep{schaye10,vandevoort12,narayanan15}.
Recently \citet{thone19} also suggested that in GRB hosts gas outflows are very common.

Observations of atomic gas in host galaxies of unusual and/or unclassified transients can therefore bring us closer to understanding the nature of these events. Similar atomic gas properties around the position of a transient to those of GRBs would suggest that the explosion mechanism is similar, i.e., an explosion of a massive star.

With this in mind, we report an analysis of gas properties in the host galaxy of the unusual and poorly-understood transient
 \object{AT 2018cow}\footnote{It was initially designated \object{ATLAS\,18qqn} by the ATLAS discovery team.}.
 The transient was discovered on 16 June 2018 by the Asteroid Terrestrial-impact Last Alert System \citep[ATLAS;][]{atlas} surveying the entire visible sky every two nights  
\citep{smartt18atel,prentice18}.  It was classified as a broad-lined type Ic \citep{izzo18atel,xu18atel}, type Ib \citep{benetti18atel},
 or interacting type Ibn \citep{fox19}
supernova and given the designation \object{SN 2018cow}. However, it is unclear whether this really was a supernova (see below).
It has been detected at (sub){\mm} \citep{deugartepostigo18atel,smith18atel,ho19} and radio \citep{dobie18atel,bright18atel,nayana18atel,margutti19} wavelengths, including very long-baseline interferometry (VLBI) from which the most precise position has been derived: RA (J2000) = 16:16:00.2243, Dec. (J2000) = +22:16:04.893 with a $\sim1\,$mas uncertainty   \citep{an18atel,bietenholz18atel,horesh18atel}.  
{\atcow} had several unusual characteristics: 
high peak luminosity, blue colour/high temperature even a month after the explosion, very fast initial flux rise \citep{prentice18,perley19}, 
high decline rate, no spectral features up to four days after the explosion, very broad short-lived absorption and emission spectral features \citep{perley19},
variability of the X-ray light curve \citep{riverasandoval18},
a month-long plateau at {\mm} wavelengths (see \citealt{michalowski18grb} for another example), as well as high radio flux  \citep{ho19}.

It has been shown that it could not have been powered by radioactive decay \citep{prentice18,margutti19,perley19}.
Several models have been proposed to explain the observed properties:
a stellar collapse leading to the formation of a magnetar \citep{prentice18,margutti19},
a luminous blue variable exploding in a non-uniform circum-stellar medium \citep{riverasandoval18},
a SN from a low-mass hydrogen-rich star, a failed SN from a blue supergiant \citep{margutti19},
a tidal disruption event \citep[TDE;][]{liu18,kuin19,perley19},
a jet driven by an accreting neutron star colliding with a giant star \citep{soker19},
or a merger of white dwarfs \citep{lyutikov19}.
However, the constraints on the nature of this explosion set by the host galaxy properties have not been explored thoroughly.

{\atcow} exploded within a spiral galaxy of type Sc \citep{willett13}, \object{CGCG 137-068}, at a redshift of 
$z=0.014$ \citep{perley19}.
It has an inclination from the line of sight of $24.4^\circ$ \citep{hyperleda}.\footnote{\hyperleda.}
It has a bar and weak spiral arms \citep{perley19}. 
Its stellar mass and SFR are $1.42^{+0.17}_{-0.29}\times10^9\,\msun$ and $0.22^{+0.03}_{-0.04}\,\msunyr$, respectively \citep{perley19}.
The galaxy was claimed to be asymmetric with more near-IR emission in the south-west, i.e., in the part of the galaxy where {\atcow} exploded \citep{kuin19},
$\sim1.7$\,kpc from the galaxy centre \citep{kuin19,perley19}.

\newlength{\szerkol}
\setlength{\szerkol}{0.5\textwidth}

\begin{figure*}
\begin{center}
\begin{tabular}{cc}
\includegraphics[width=\szerkol]{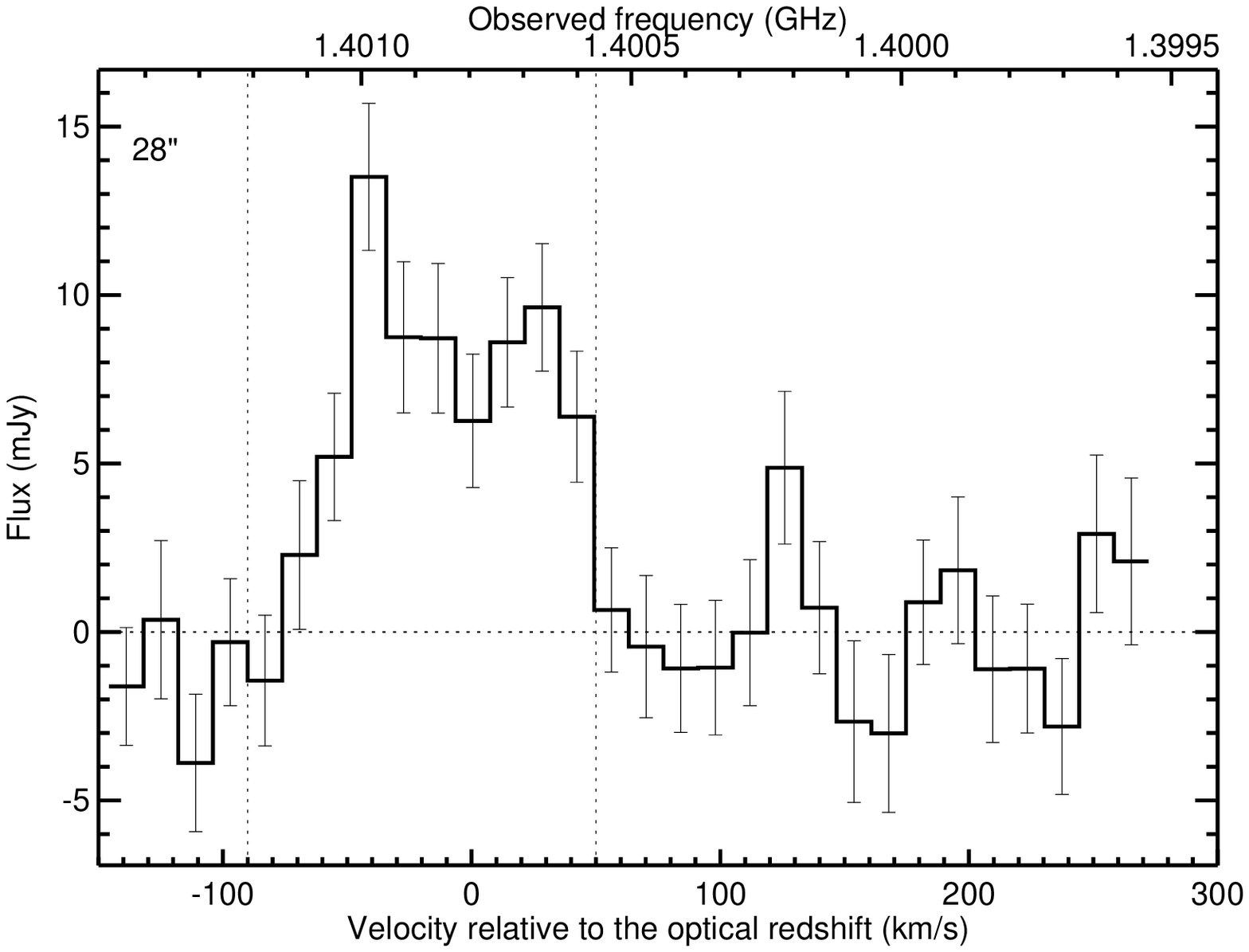} &

\includegraphics[width=\szerkol]{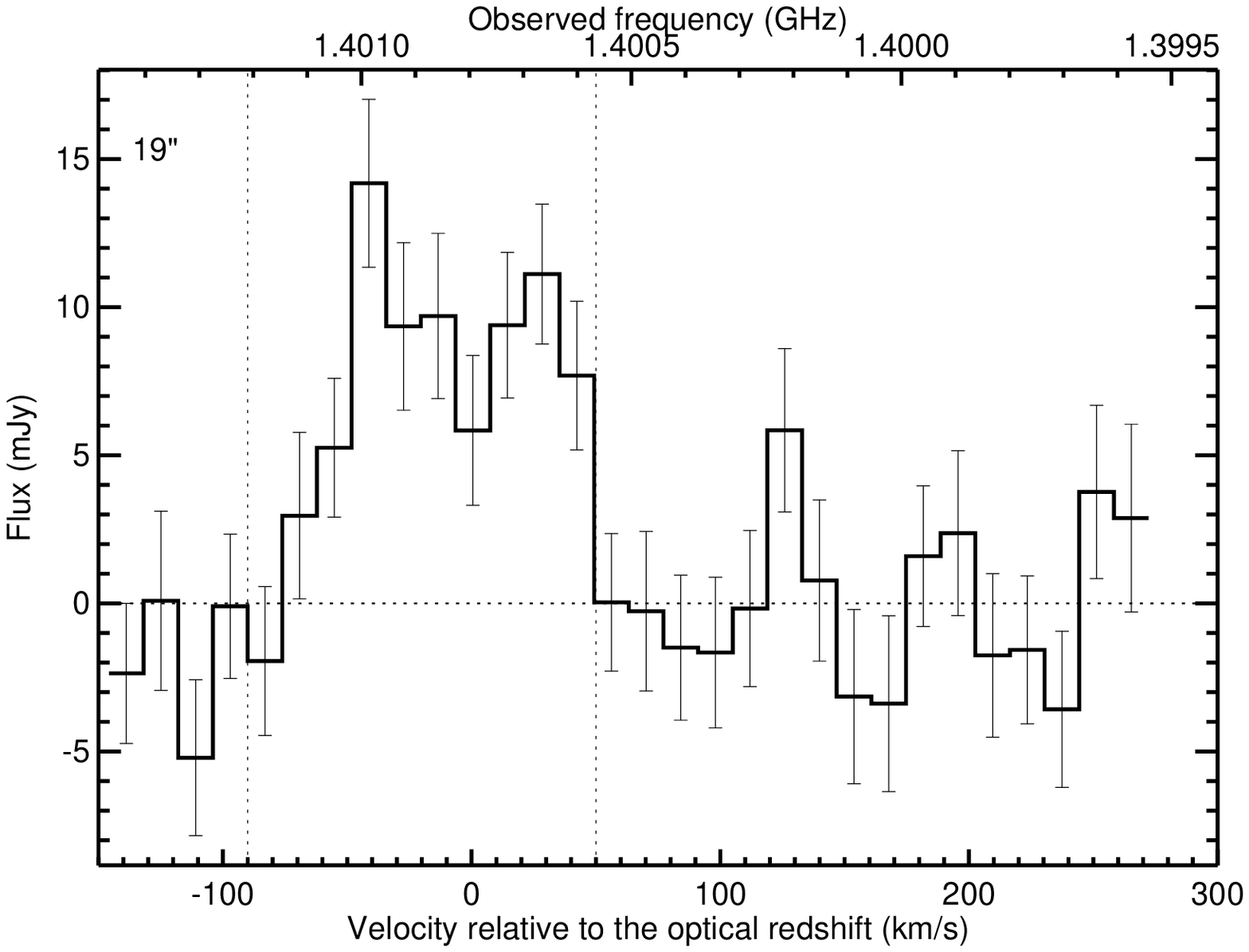} \\

\includegraphics[width=\szerkol]{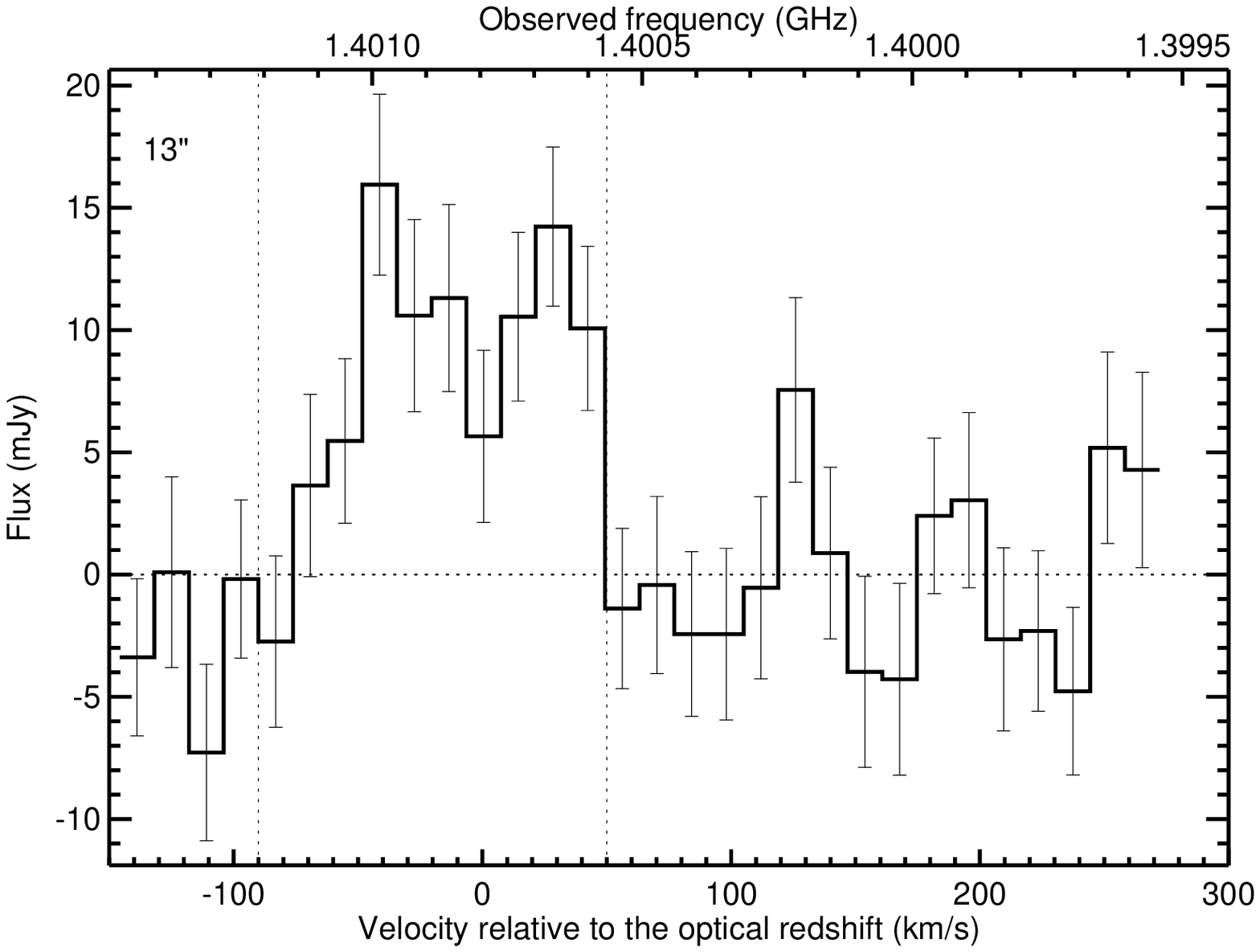} &
\includegraphics[width=\szerkol]{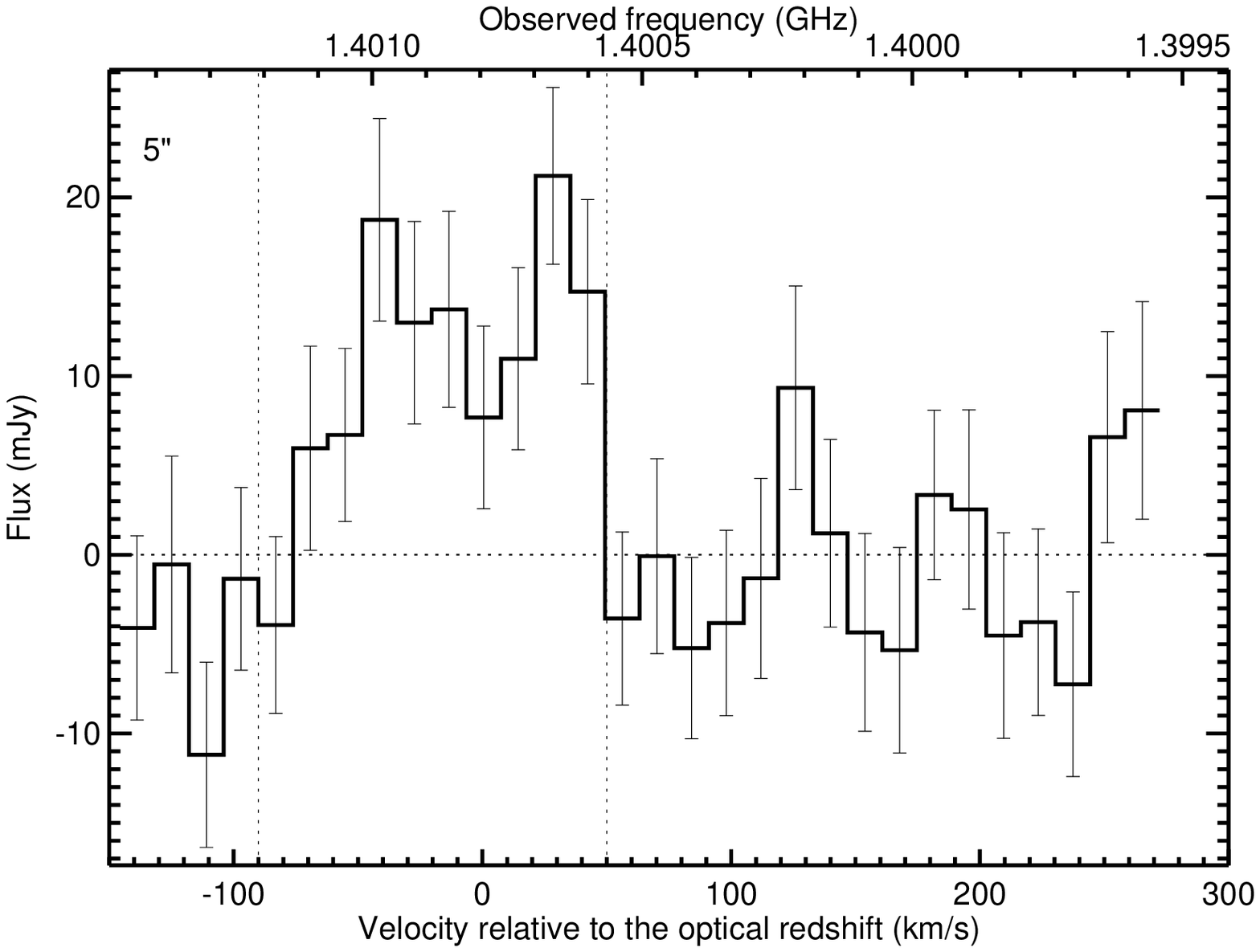} 
\end{tabular}
\end{center}
\caption{{\hi} spectra of {\host} extracted over the entire galaxy within an aperture of $45\arcsec$ radius (solid histogram) derived from the data cubes with resolutions 
as marked on the panels.
The {\it dotted lines} denote the velocity range over which the total {\hi} was estimated.
}
\label{fig:hispec}
\end{figure*}

The objectives of this paper are: {\it i)} to provide a resolved measurement of the atomic gas properties of the host galaxy of {\atcow} in order to learn about its nature, and {\it ii)} to test whether these properties are consistent with a recent inflow of atomic gas from the intergalactic medium.

We use a cosmological model with $H_0=70$ km s$^{-1}$ Mpc$^{-1}$,  $\Omega_\Lambda=0.7$, and $\Omega_m=0.3$, implying that {\atcow}, at $z= 0.014$, is at a luminosity distance of 60.6 Mpc and $1\arcsec$ corresponds to 286 pc at its redshift. We also assume the 
\citet{chabrier03} 
initial mass function (IMF).

\section{Data}
\label{sec:data}

\setlength{\szerkol}{0.23\textwidth} 

\begin{figure*}
\begin{center}
\begin{tabular}{cccc}
\includegraphics[width=\szerkol]{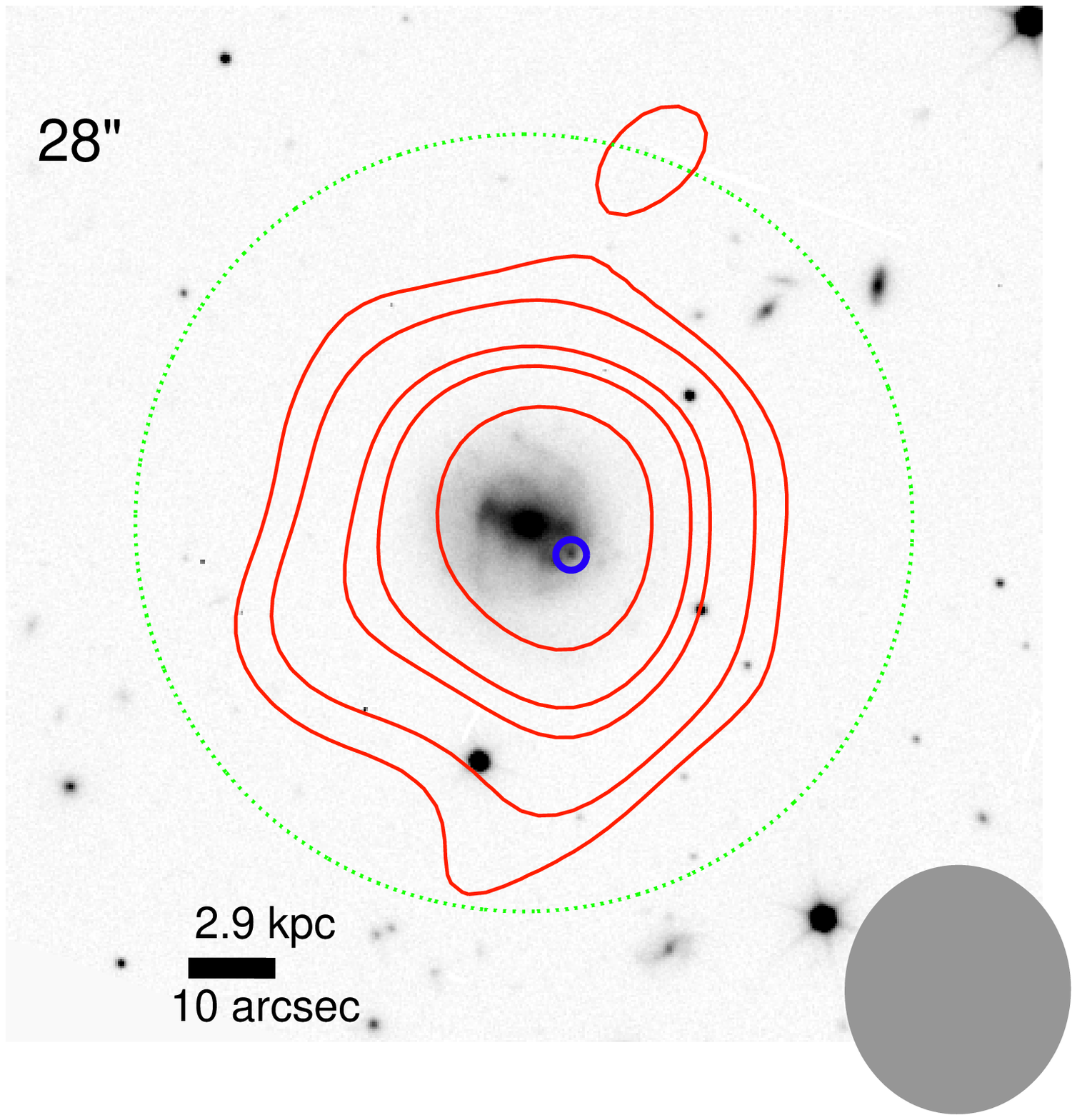} &
\includegraphics[width=\szerkol]{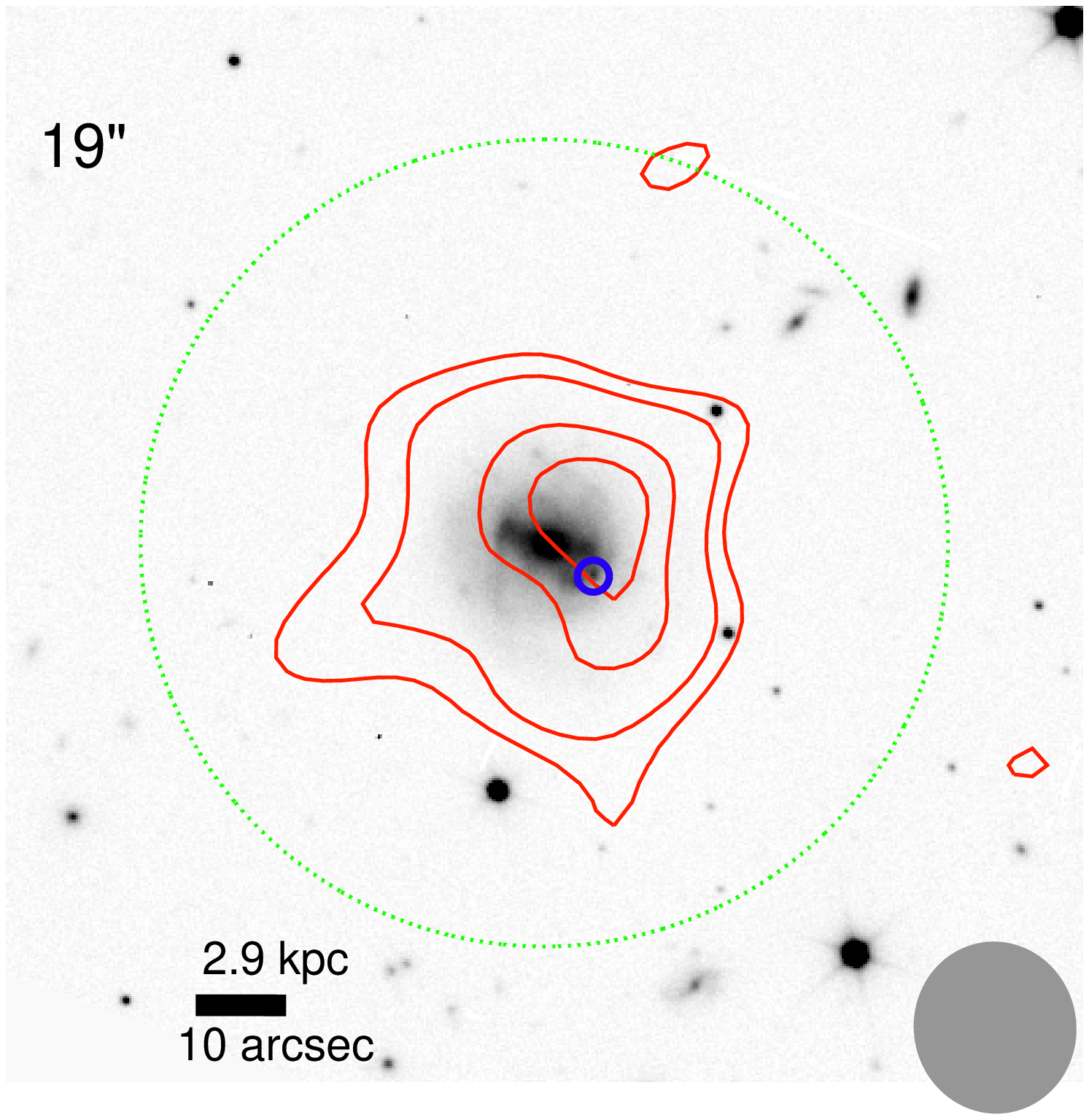} &
\includegraphics[width=\szerkol]{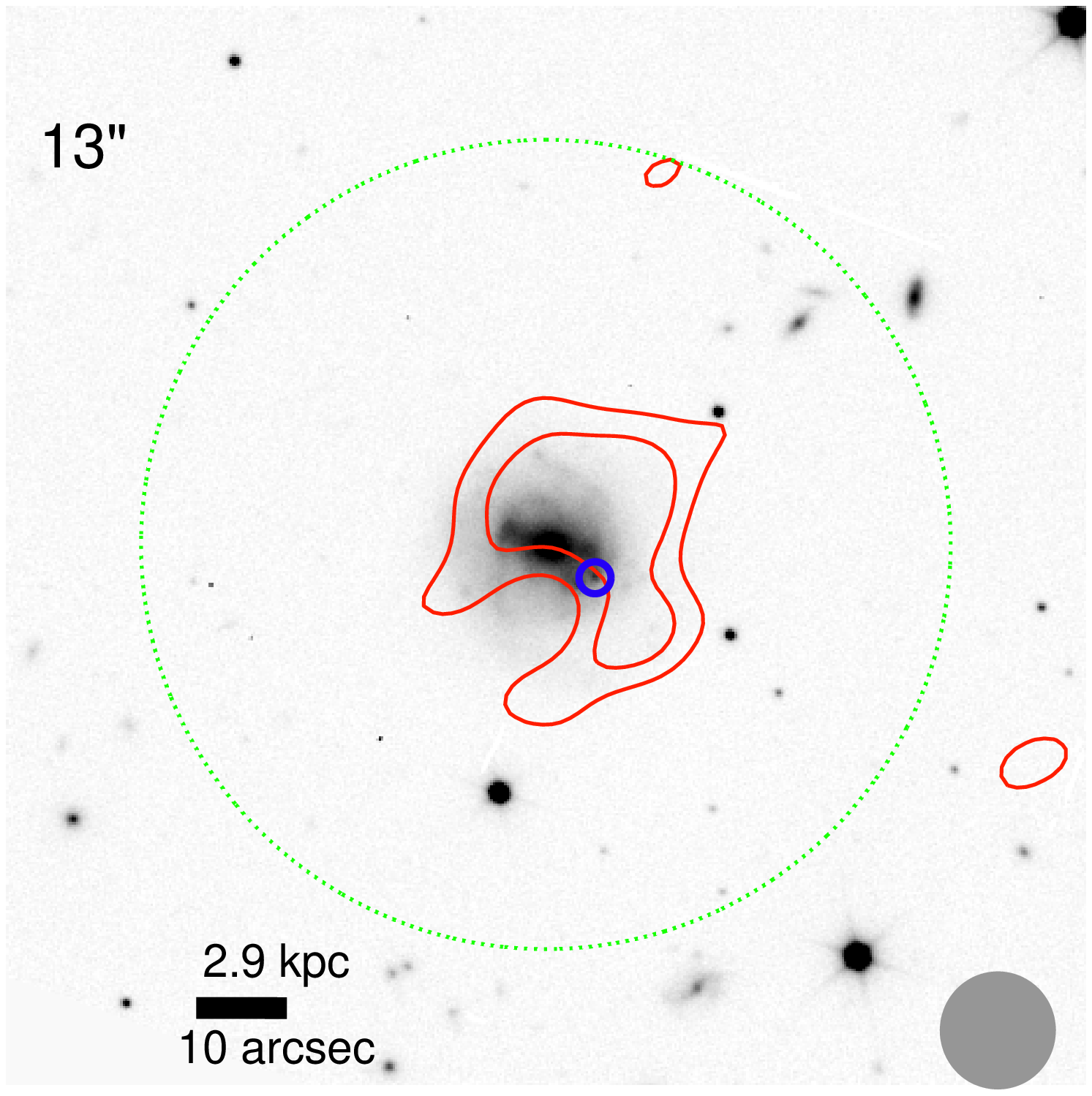} &
\includegraphics[width=\szerkol]{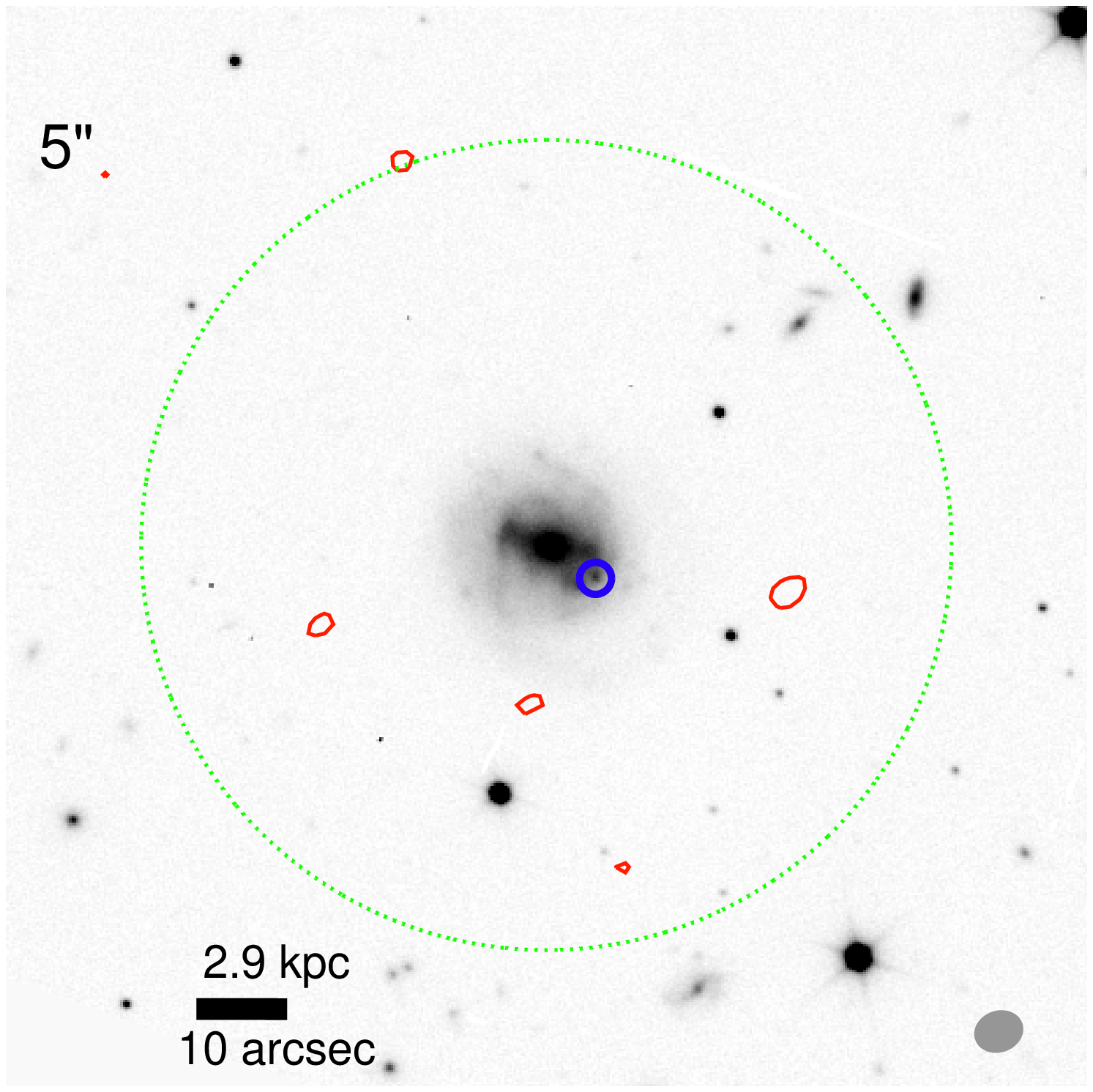} \\
\includegraphics[width=\szerkol]{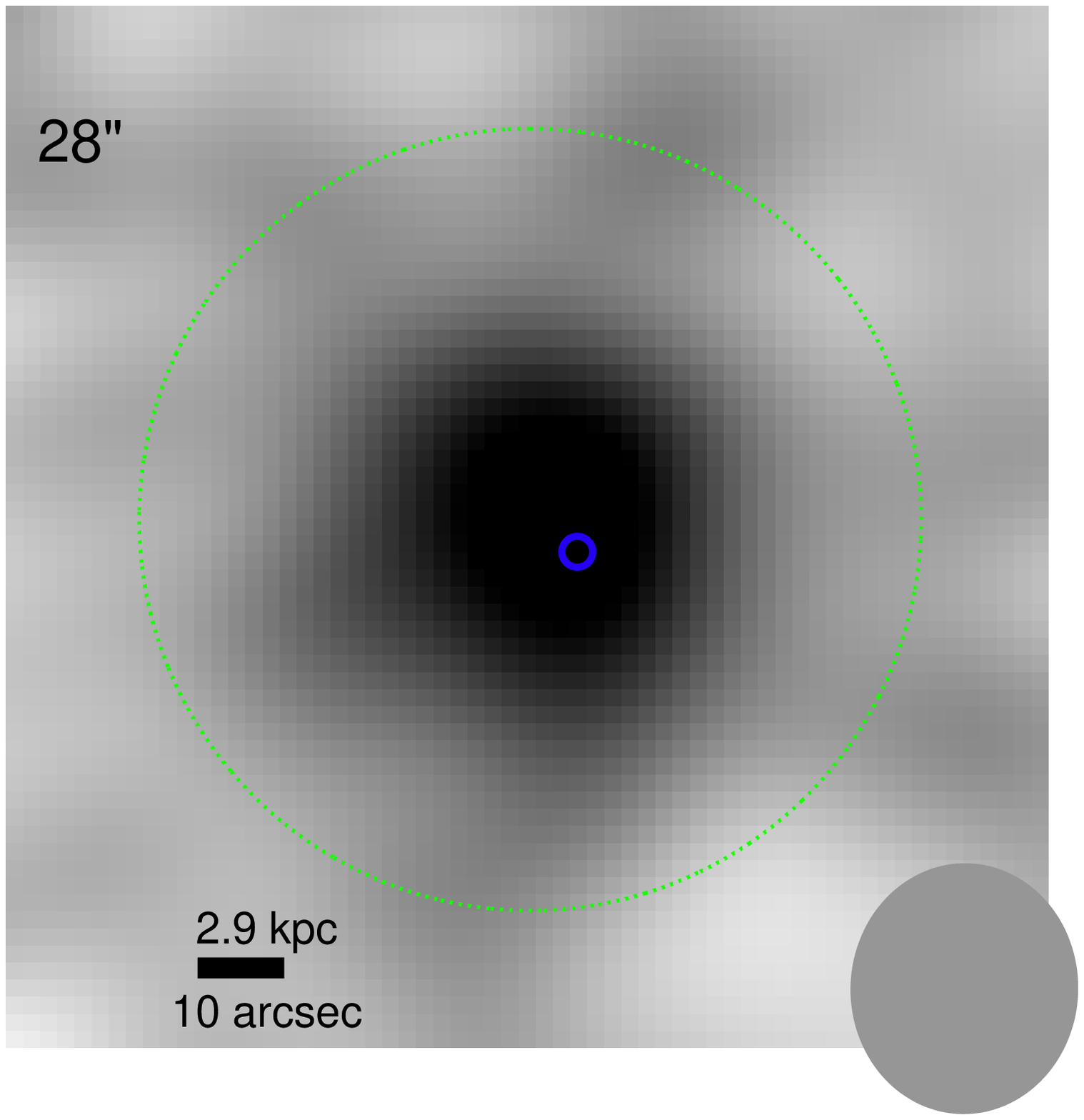} &
\includegraphics[width=\szerkol]{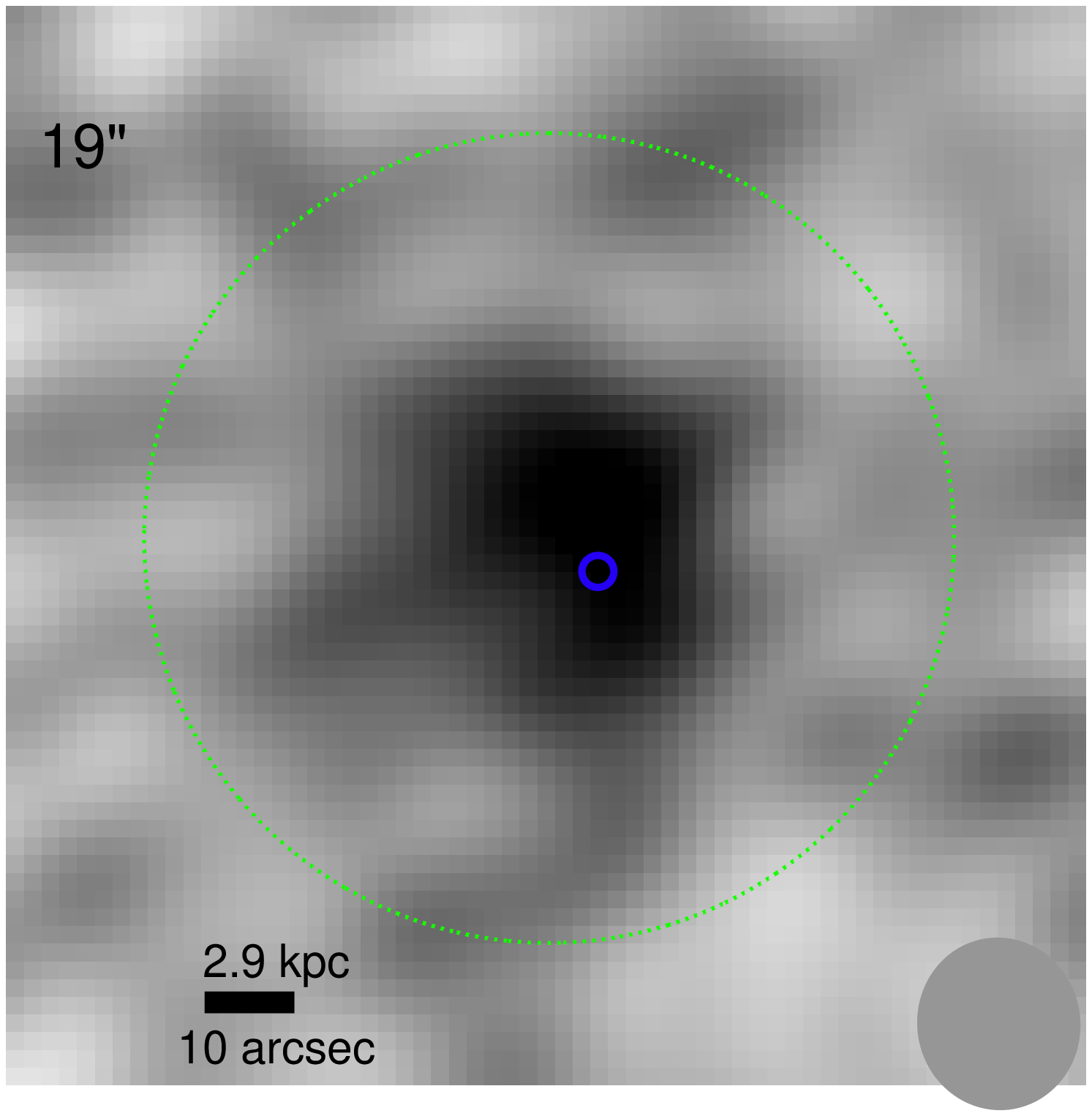} &
\includegraphics[width=\szerkol]{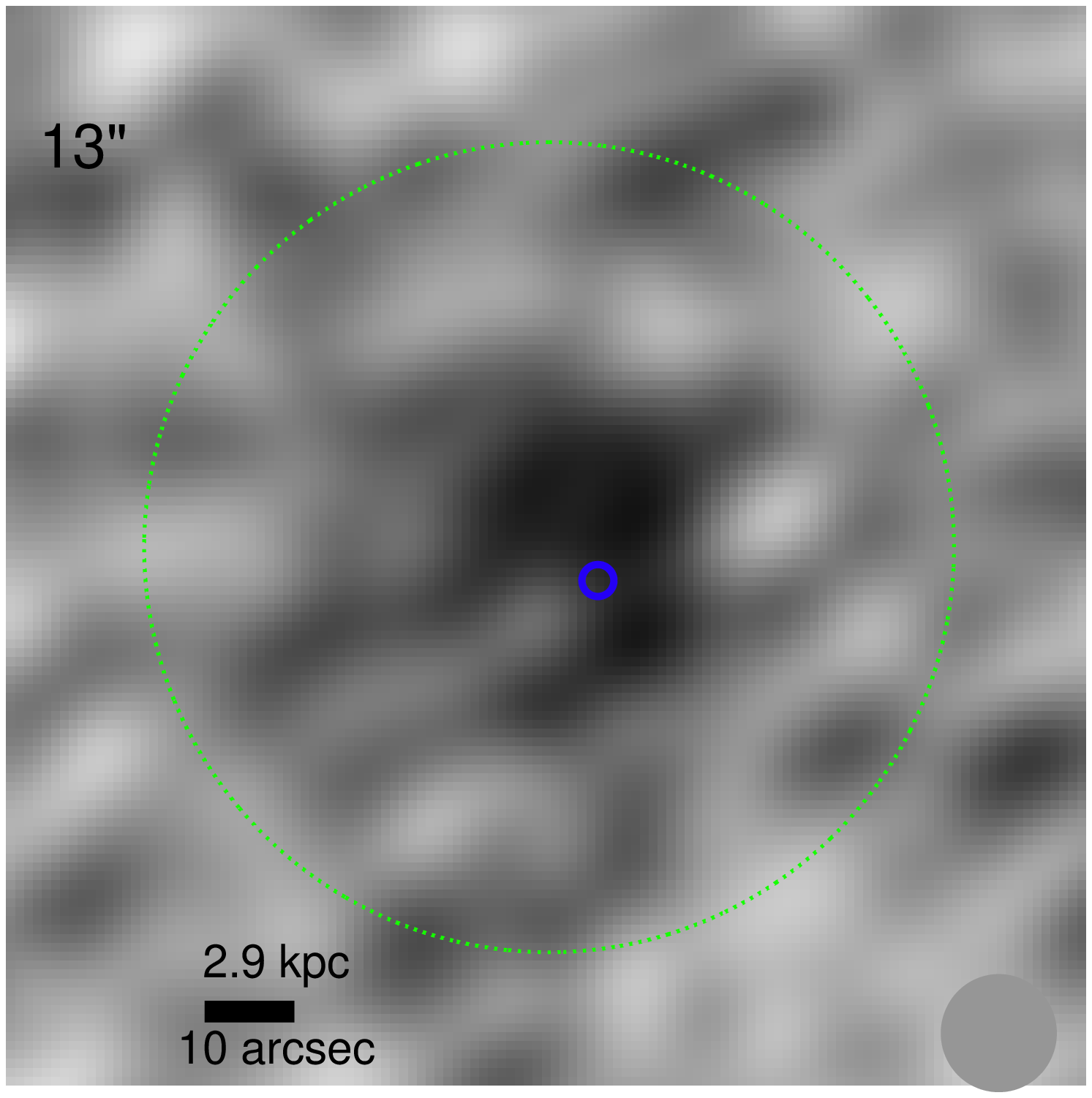} &
\includegraphics[width=\szerkol]{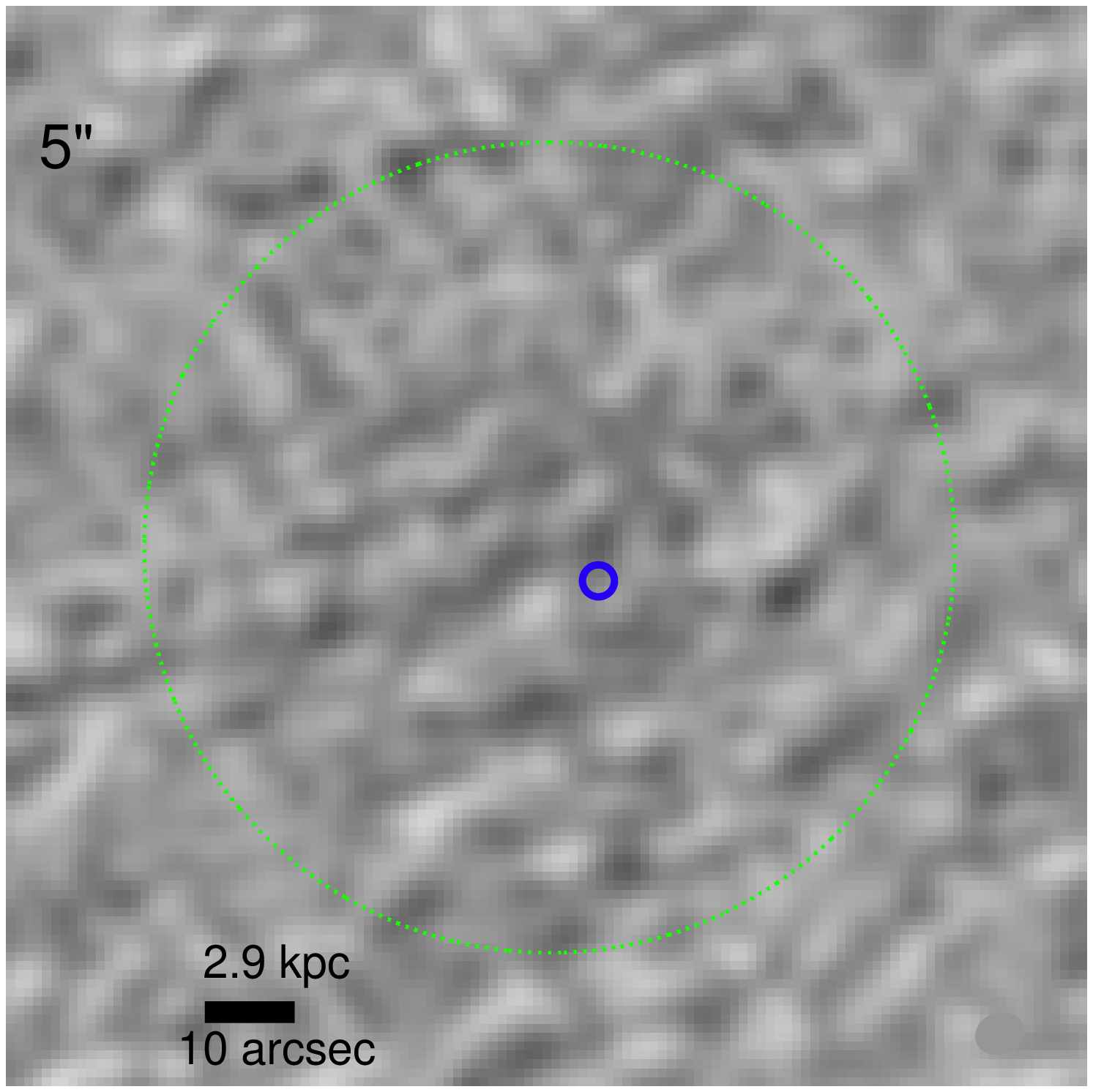} \\
\includegraphics[width=\szerkol]{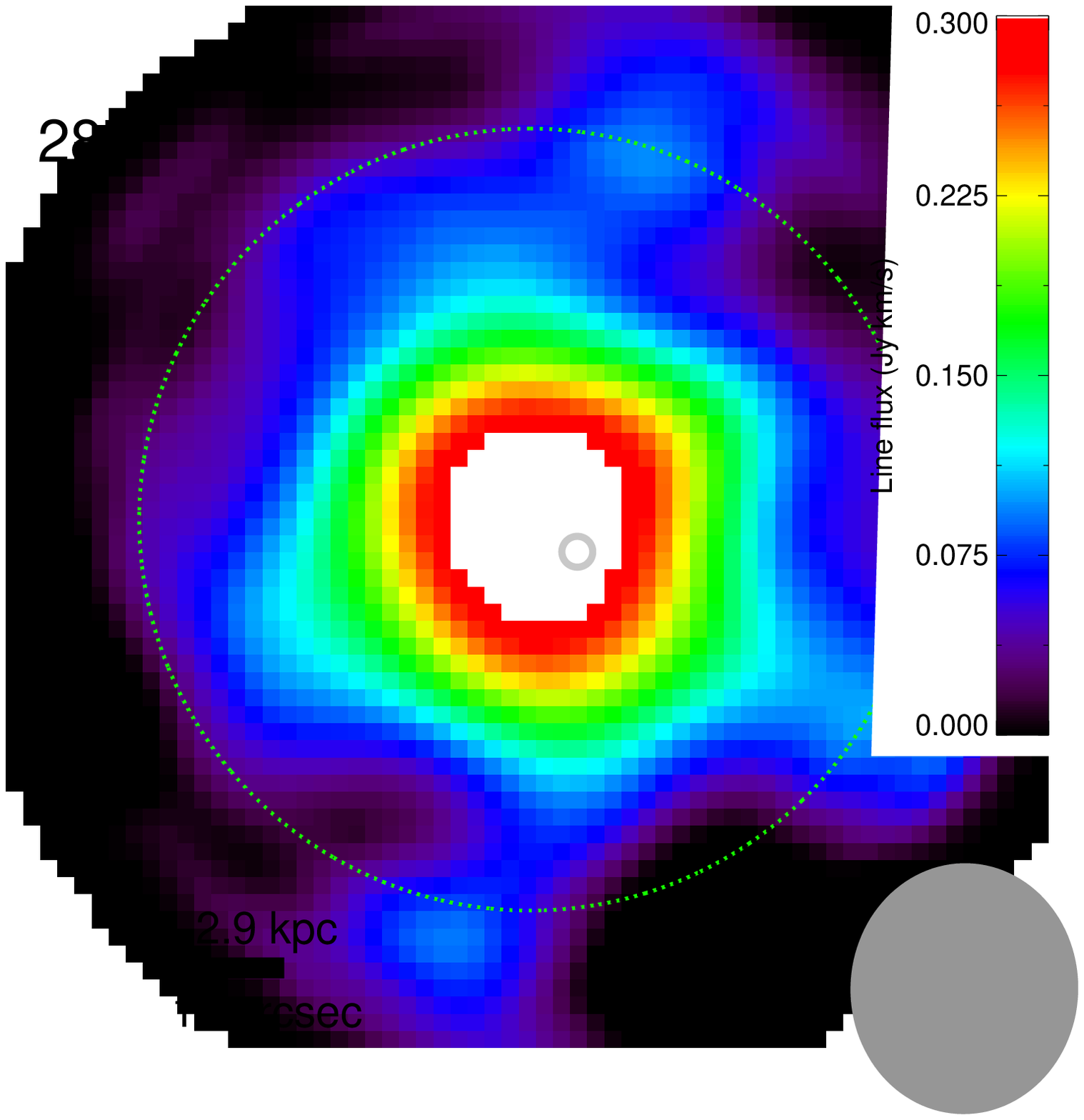} &
\includegraphics[width=\szerkol]{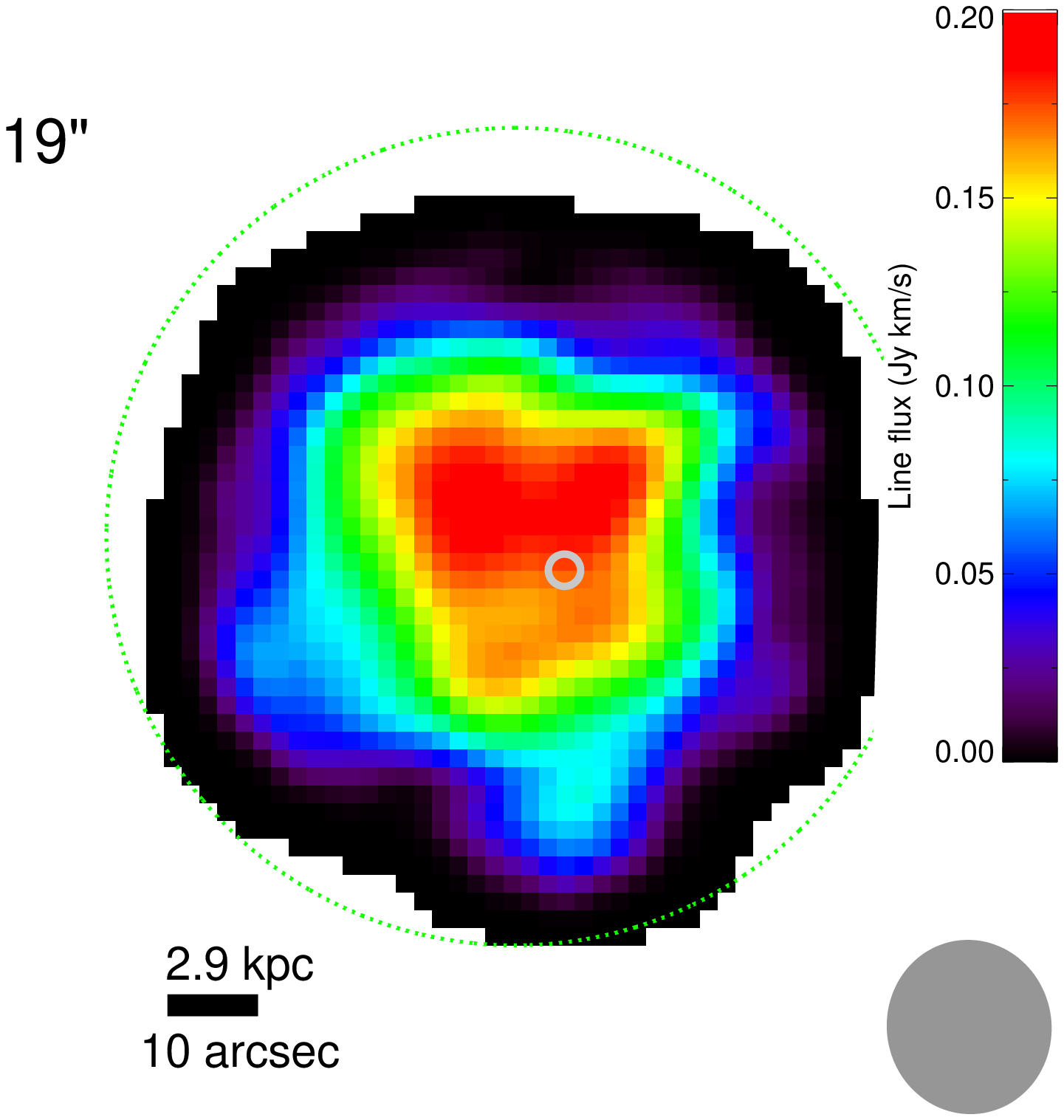} &
\includegraphics[width=\szerkol]{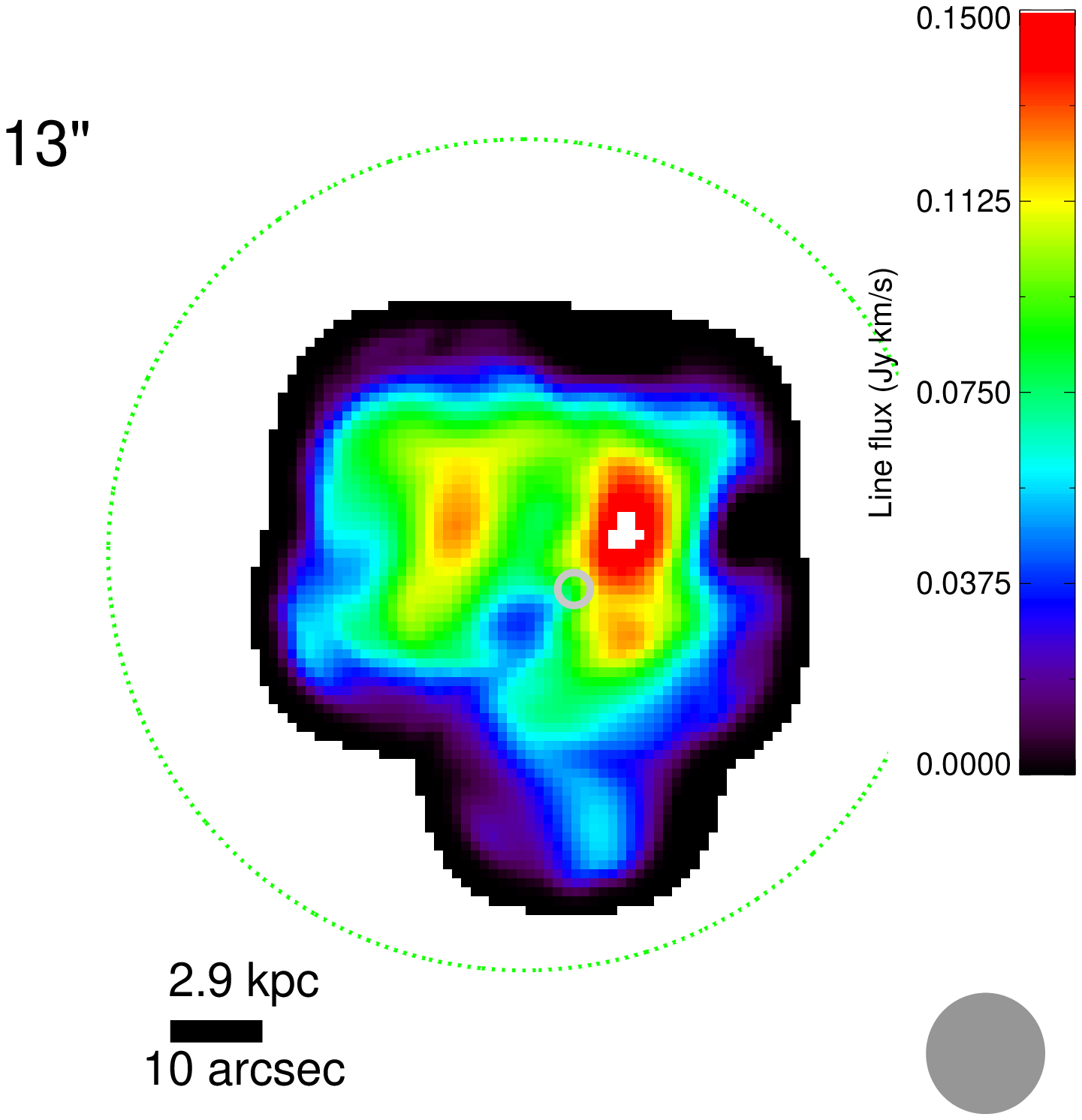} &
\includegraphics[width=\szerkol]{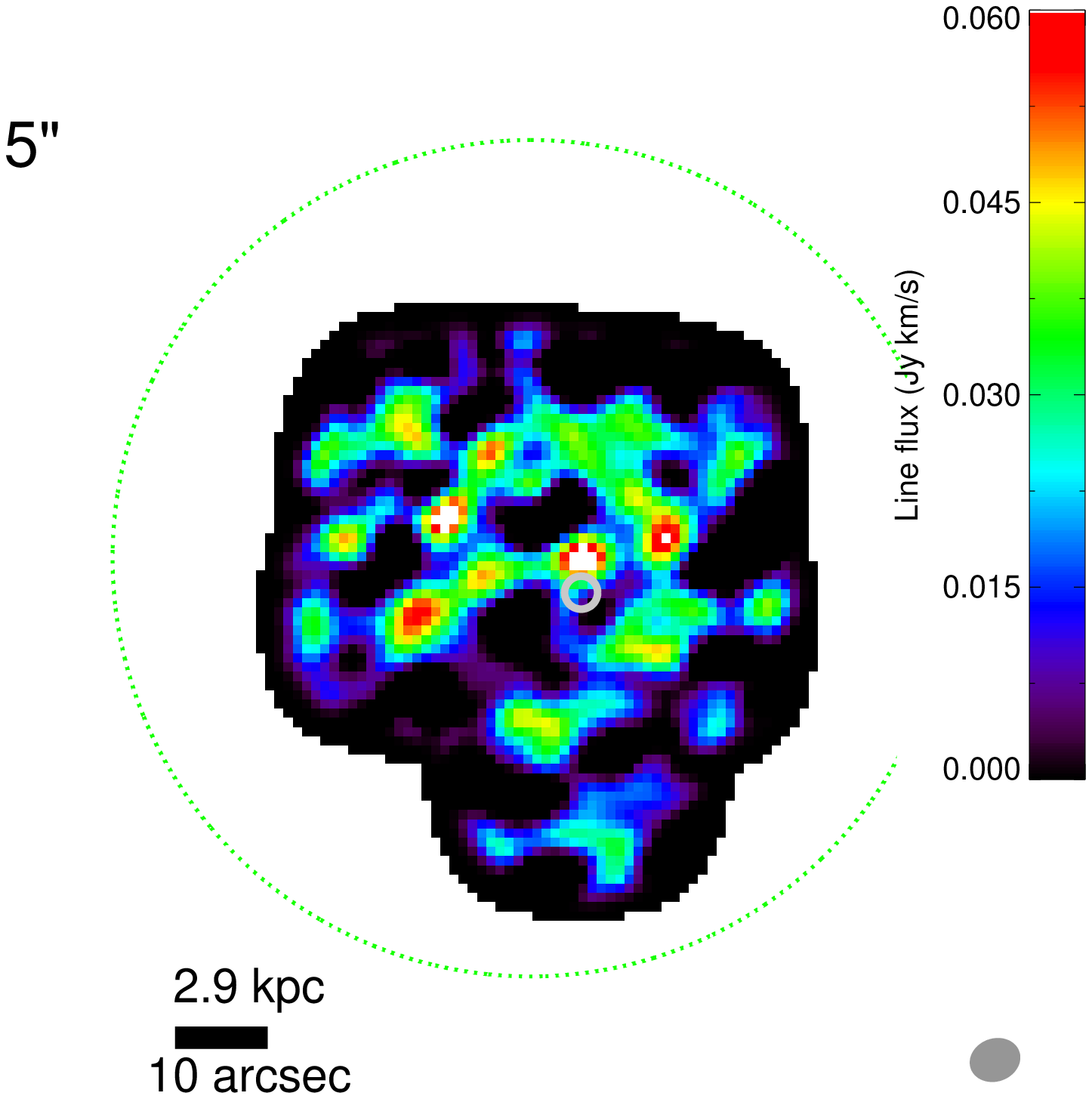} \\
\includegraphics[width=\szerkol]{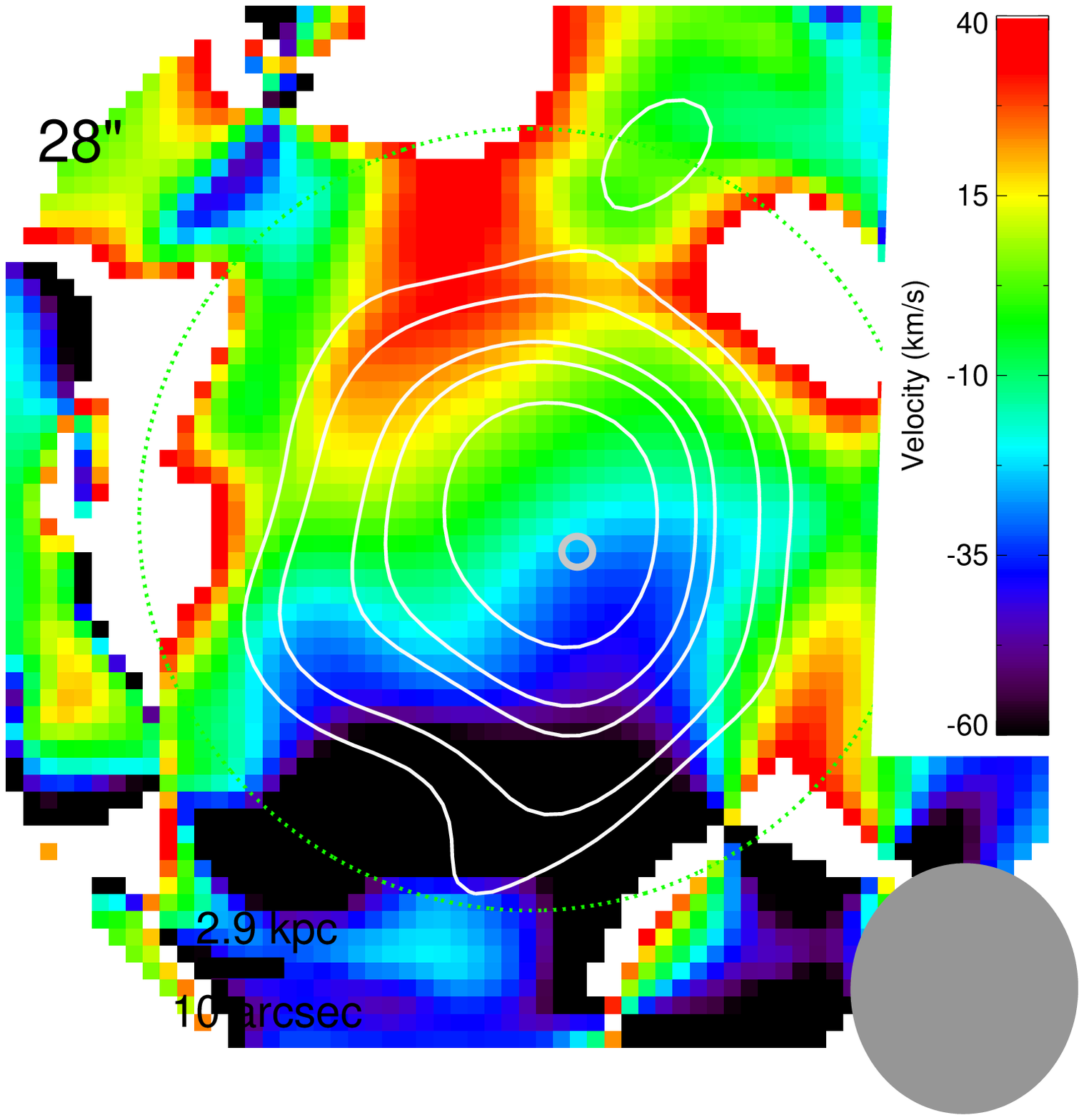} &
\includegraphics[width=\szerkol]{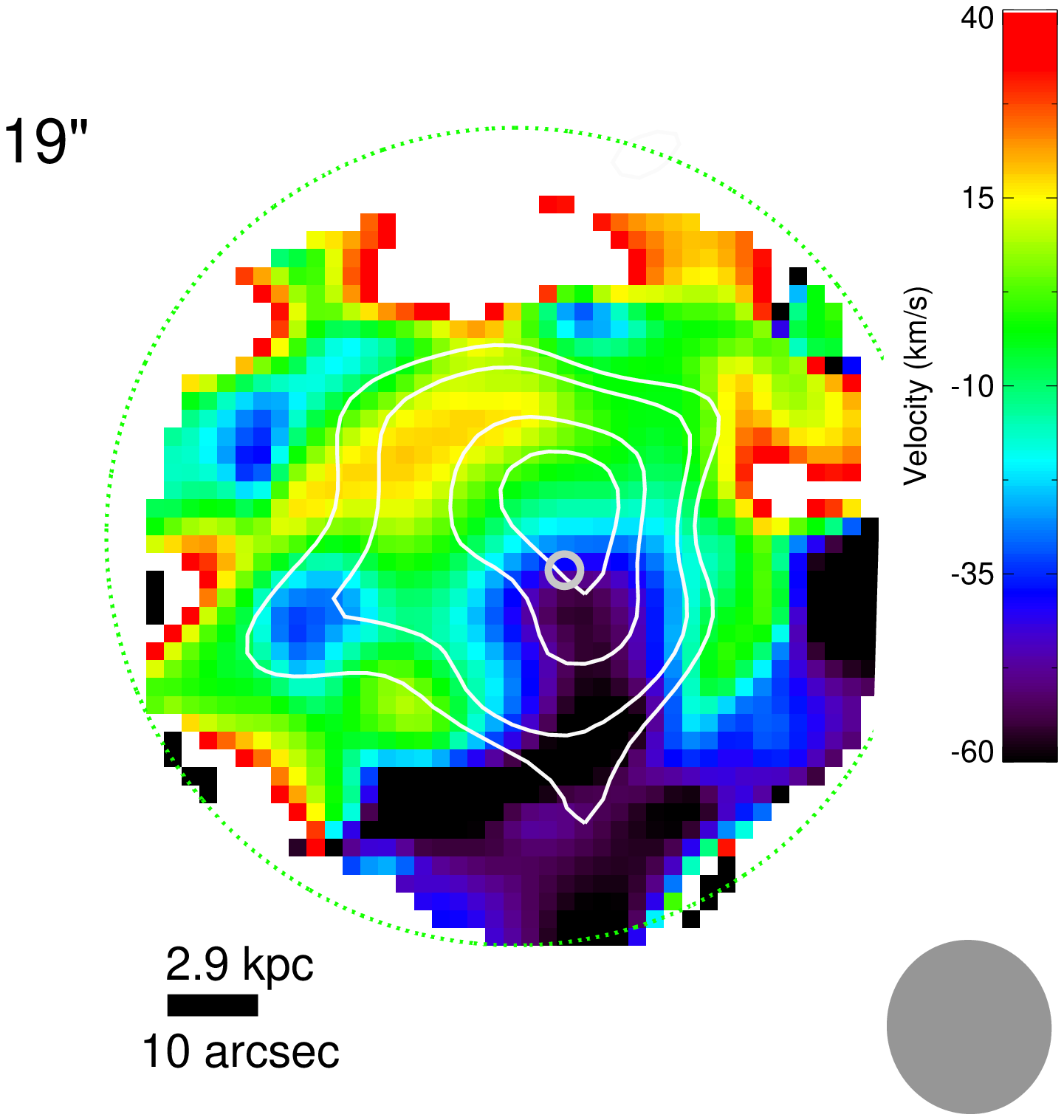} &
\includegraphics[width=\szerkol]{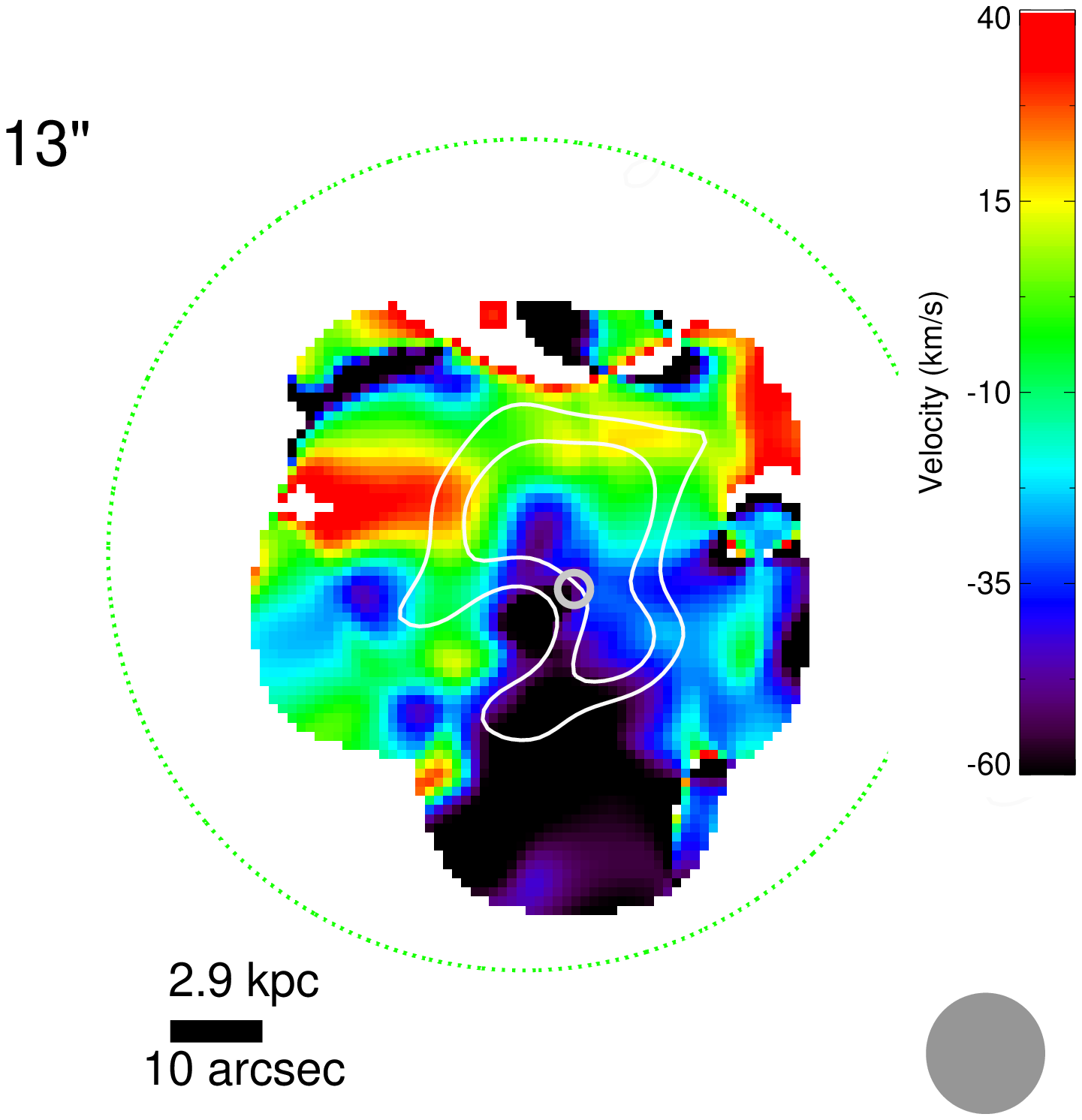} &
\includegraphics[width=\szerkol]{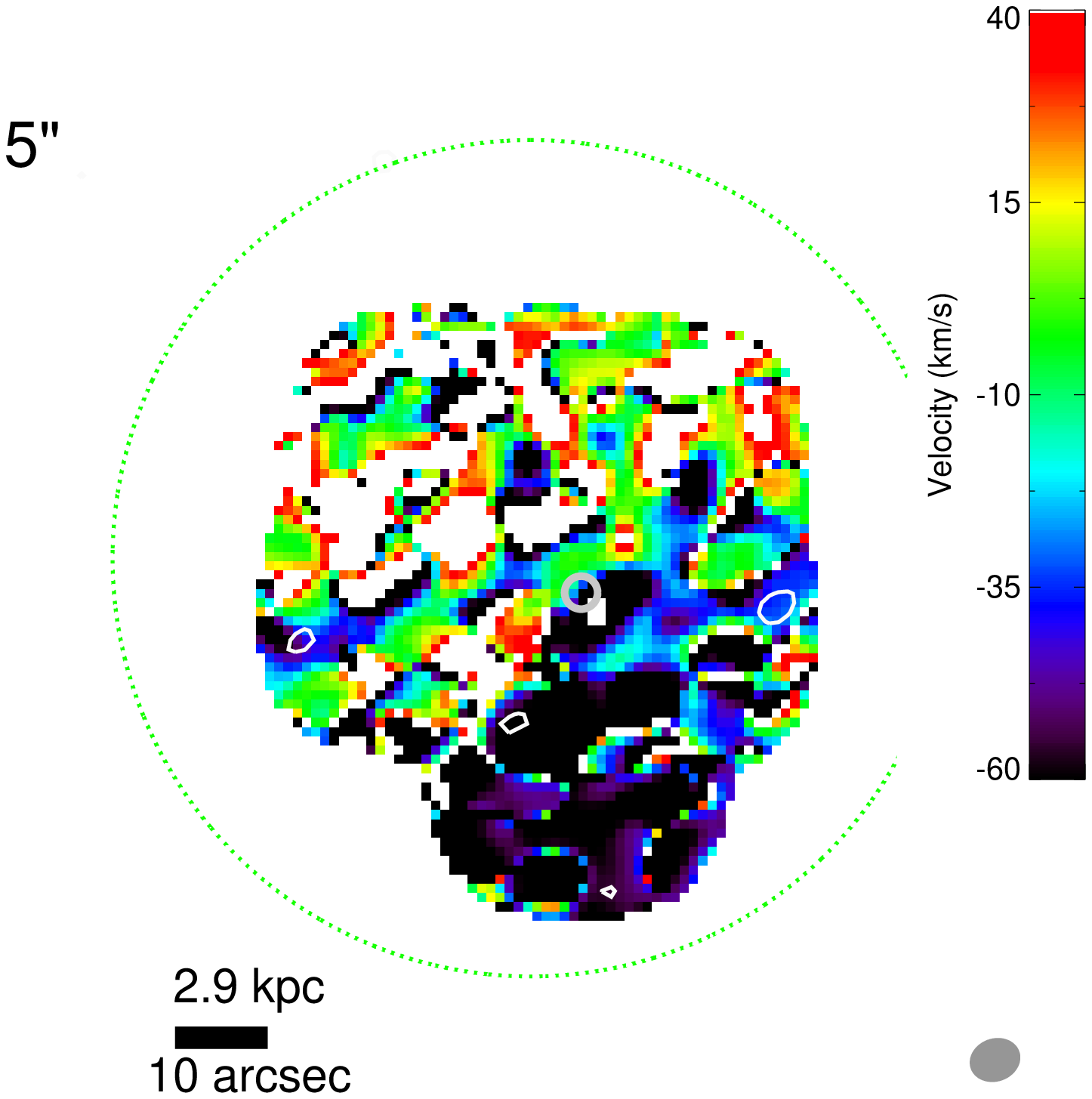} \\
\end{tabular}
\end{center}
\caption{{\it Top}: {\hi} contours (red;  collapsed {\hi} cube) of {\host} overlayed on the Gran Telescopio Canarias optical $i^\prime$-band image (Kann et al., in prep.). 
The contours are 3, 4, 6, 7, and 9$\sigma$, where $\sigma=0.031$, $0.029$, $0.027$, and $0.019\,\mbox{Jy beam}^{-1}\,\kms$ for the data at the resolution of 28\arcsec, 19{\arcsec}, 13{\arcsec}, and 5\arcsec, respectively (corresponding to a neutral hydrogen column density of $\sim 0.5$, $0.9$, $1.8$, and $8.4\times10^{20}\,\mbox{cm}^2$, respectively).   
{\it Second row}: the {\hi} data cube collapsed within the dotted lines given in Fig.~\ref{fig:hispec}.
{\it Third row}: 
The zeroth moment map (integrated emission) of the {\hi} line.
{\it Bottom}:
The first moment map (velocity field) of the {\hi} line with the same contours as in the top panel. The velocities are relative to the systemic velocity of $4197\,\kms$ derived from the optical spectrum \citep{perley19}.
{\it Columns} are for the resolution 
as marked on the panels.
The VLBI position of {\atcow} is indicated by the {\it blue} or {\it grey circles}.
The {\it green dotted circle} has a radius of 45{\arcsec} and corresponds to the aperture within which the total {\hi} emission was measured.
The beam size of the {\hi} data is shown as the {\it grey ellipses}.
The images are $120\arcsec\times120\arcsec$ and the scale is indicated by the ruler. North is up and East is to the left.
}
\label{fig:all}
\end{figure*}

On 8 and 9 February 2019 the field of {\atcow} was observed for 14 hrs with the Giant Metrewave Radio Telescope (GMRT)\footnote{Project no.~35\_021, PI: M.~Micha{\l}owski}. For calibration of the flux and the bandpass 3C286 was observed for 15 min at the start and the end of the run. For the phase calibration 1609+266  was observed every 40 min. The correlator was setup with 33 Mhz bandwidth and 512 channels centred around 1400 MHz.

After the submission of this paper, additional GMRT data were reported by \citet{roychowdhury19}. Hence, for the {\hi} analysis we also included these archival data\footnote{Project no.~DDTC022, PI: M.~Arabsalmani}. As part of that program 7 hrs of data were obtained on 27 August 2018. The channel width was half as wide as for our observations. The same calibrators were observed.

The data were reduced with a range of data reduction packages. We downloaded the \texttt{FITS} files with the raw data from the GMRT archive. These \texttt{FITS} files were then loaded into the Common Astronomy Software Applications ({\sc CASA}) package \citep{casa} with the {\sc importgmrt} task without applying the online flags. Further data reduction was done with the {\sc meerkathi}\footnote{\url{https://github.com/ska-sa/meerkathi}, a private repository for the time of development} pipeline which is being developed for {\hi} data reduction of MeerKAT data. The pipeline is setup in a modular fashion using the platform-independent radio interferometry scripting framework {\sc stimela}\footnote{\url{https://github.com/SpheMakh/Stimela/wiki}}. In practice this means that the calibrator data  are initially flagged with {\sc AOflagger} \citep{offringa10} and calibrated and transferred to the target with {\sc CASA}. For the data in this paper the phase calibrator was used as a bandpass calibrator, as the GMRT bandpass clearly fluctuated over the time of the observations. As the phase calibrator was bright (4.8 Jy), this leads to an improved bandpass calibration over the course of the observations.

After the initial calibration the target was split out of the measurement set, further flagged with {\sc AOflagger}, imaged with {\sc WSclean} \citep{offringa14} in Stokes I, then the sources in the field were extracted and modeled with pyBDSF\footnote{\url{https://github.com/lofar-astron/PyBDSF}}, after which this model was used in Cubical\footnote{\url{https://github.com/ratt-ru/CubiCal/}} \citep{kenyon18} for the self-calibration. This step was repeated until a phase-only self-calibration no longer improved the extracted models.

After the calibration, the three separate days were mapped on to the same channel grid with the {\sc CASA} task {\sc mstransform} and the modelled continuum was subtracted from the data. Any residual continuum was subtracted with {\sc uvlin}. At this stage the data were also doppler-corrected and projected onto a barycentric velocity frame. 

The visibilities were weighted according to a Briggs weighting scheme with Robust = 0.0 and uvtapers of 4, 6, 8 and 20 k$\lambda$ were applied to attain cubes with varying spatial resolution. The cubes were inverted and cleaned with the {\sc CASA} task {\sc TCLEAN}. The cleaning was performed in an iterative process where we first clean the full cube to a 10$\sigma$ threshold then create a mask with {\sc SoFiA} \citep{serra15}, and then clean within this mask to 0.5$\sigma$. This last step was done outside the pipeline as currently it can not deal with the frequency increments of opposite sign in the different datasets.

The final cubes have a resolution of  
$\textnormal{FWHM}=28\farcs9\times26\farcs2$, $19\farcs2\times18\farcs1$, $13\farcs12\times12\farcs9$, and $5\farcs5\times4\farcs6$
and a channel width of 65.1 kHz. The frequency axis was converted into a velocity axis using the relativistic definition which results in a channel width of 13.9 km s$^{-1}$ with an error of $\sim$ 0.01 km s$^{-1}$ on the outermost channels of the cube. 

For our data from February 2019 (excluding those from August 2018, due to the variability of {\atcow}) we also imaged together all channels of the entire 33\,MHz bandwidth (before continuum subtraction) to produce a continuum image at an observed frequency of $1.397667$\,GHz. The beam size is $2.0\arcsec\times1.8\arcsec$ and the noise is $17.5\,\mu$Jy\,beam$^{-1}$. The {\hi} line spans $\sim0.6$\,MHz, so it should not affect this  continuum image based on the 33\,MHz bandwidth. Indeed, when we exclude the channels with the line emission we obtain an almost identical map.

In order to correct the astrometry we identified 16 sources in the Faint Images of the Radio Sky at Twenty-Centimeters (FIRST) survey \citep{first,first2} that are point-like in our continuum map. On average these sources were found to be shifted on our map with respect to the FIRST position  by $(+1.30\pm0.14)\arcsec$ in right ascension and  $(+0.51\pm0.14)\arcsec$ in declination. We shifted our continuum and {\hi} maps by this offset. This has very little effect on {\hi} maps, as their beam sizes are much larger. This offset also implies that the positional uncertainty in our continuum map is $0.14\arcsec$ in both directions.

\begin{table*}
\small
\centering
\caption{{\hi} properties of {\host}.\label{tab:mhi}}
\medskip
\begin{tabular}{lcccccc}
\hline
Beam & \zhi & $W_{50}$ & $W_{10}$ & $F_{\rm int}$   &  $\log(\lphi)$          & $\log(\mhi)$ \\
 ($\arcsec$)   &      & (\kms)  & (\kms) & (Jy km s$^{-1}$) & ($\mbox{K km s}^{-1} \mbox{ pc}^2$) & (${\rm M}_\odot$) \\
(1)    & (2)     & (3)  & (4)           & (5) & (6) & (7) \\
\hline
28 & $0.013972 \pm        0.000014$ & $50 \pm 30$ & $127 \pm 9$ & $0.94 \pm 0.09$ & $10.738 \pm 0.038$ & $8.909 \pm 0.038$  \\
19 & $0.013976 \pm        0.000014$ & $48 \pm 22$ & $127 \pm 18$ & $1.02 \pm 0.12$ & $10.772 \pm 0.048$ & $8.944 \pm 0.048$  \\
13 & $0.013983 \pm        0.000017$ & $47 \pm 22$ & $126 \pm 26$ & $1.17 \pm 0.15$ & $10.832 \pm 0.051$ & $9.004 \pm 0.051$  \\
5 & $0.013988 \pm        0.000022$ & $33 \pm 14$ & $126 \pm 34$ & $1.49 \pm 0.23$ & $10.939 \pm 0.063$ & $9.111 \pm 0.063$  \\
\hline
\end{tabular}
\tablefoot{
(1) Beam size of the {\hi} cube (the global estimates are the most reliable for the coarsest resolution). (2) Redshift determined from the emission-weighted frequency of the {\hi} line. (3) {\hi} linewidth at the 50\% of the maximum. (4) Width at the 10\% of the maximum. (5) Integrated flux within the dotted lines on Fig.~\ref{fig:hispec}. (6) {\hi} line luminosity using equation 3 in \citet{solomon97}. (7) Neutral hydrogen mass  using equation 2 in \citet{devereux90}.
}
\end{table*}

\begin{table}
\small
\centering
\caption{Properties of the continuum $1.397667$\,GHz sources within {\host}.\label{tab:cont}}
\medskip
\begin{tabular}{cccc}
\hline
RA & Dec & $F_{\rm 1.4\,GHz}$ & SFR$_{\rm radio}$ \\
 (h m s)   & (d m s)      & (mJy) & ($\msunyr$) \\
\hline
 16 16 00.209 & +22 16 04.78 & $1.239 \pm 0.018$ & $<0.565$ \\
 16 16 00.729 & +22 16 09.52 & $0.101 \pm 0.018$ & $0.081 \pm 0.011$ \\
\hline
\end{tabular}
\tablefoot{
The first object corresponds to {\atcow}. We treated its radio SFR estimate \citep[using the conversion of][]{bell03} as an upper limit because {\atcow} has a significant contribution to the radio flux. The mean time of the observations is 2019-02-08-17.41667 UT (237.28470\,days after the optical discovery).
}
\end{table}

\begin{figure}
\includegraphics[width=0.5\textwidth]{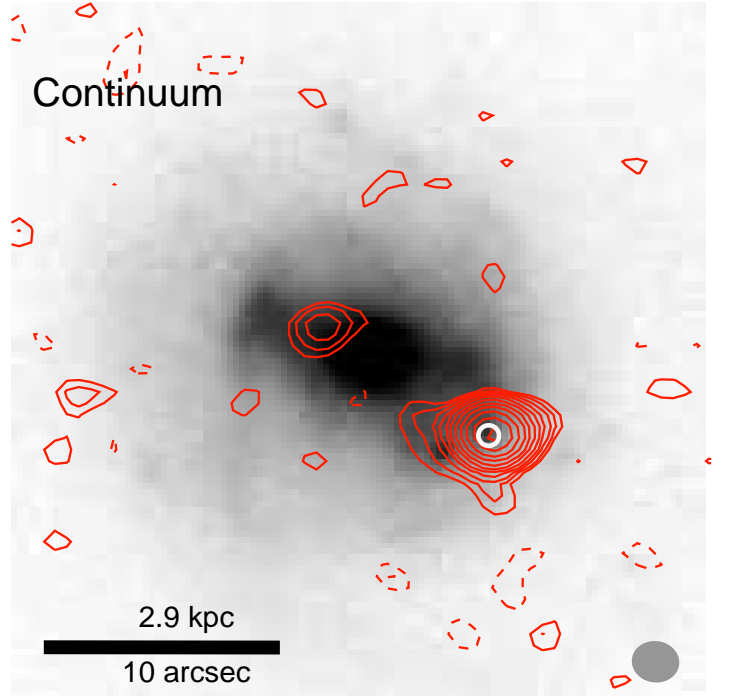}
\caption{Continuum GMRT 1.4\,GHz contours (red) of {\host}  on the Gran Telescopio Canarias optical $i^\prime$-band image of the galaxy (Kann et al., in prep.). The lowest contour is at $2\sigma$ ($\sigma=17\,\mu$Jy\,beam$^{-1}$) and the steps are in factors of $\sqrt{2}$.
The VLBI position of {\atcow} is indicated by the white circle. We detected the emission of {\atcow} at this position and an additional object in the north-eastern part of the galaxy. The beam size of the radio data is shown as the grey circle.  
The image is $30\arcsec\times30\arcsec$ and the scale is indicated by the ruler.
North is up and East is to the left.
}
\label{fig:contmap}
\end{figure}

\section{Results}
\label{sec:res}

\begin{figure*}
    \includegraphics[width=\textwidth]{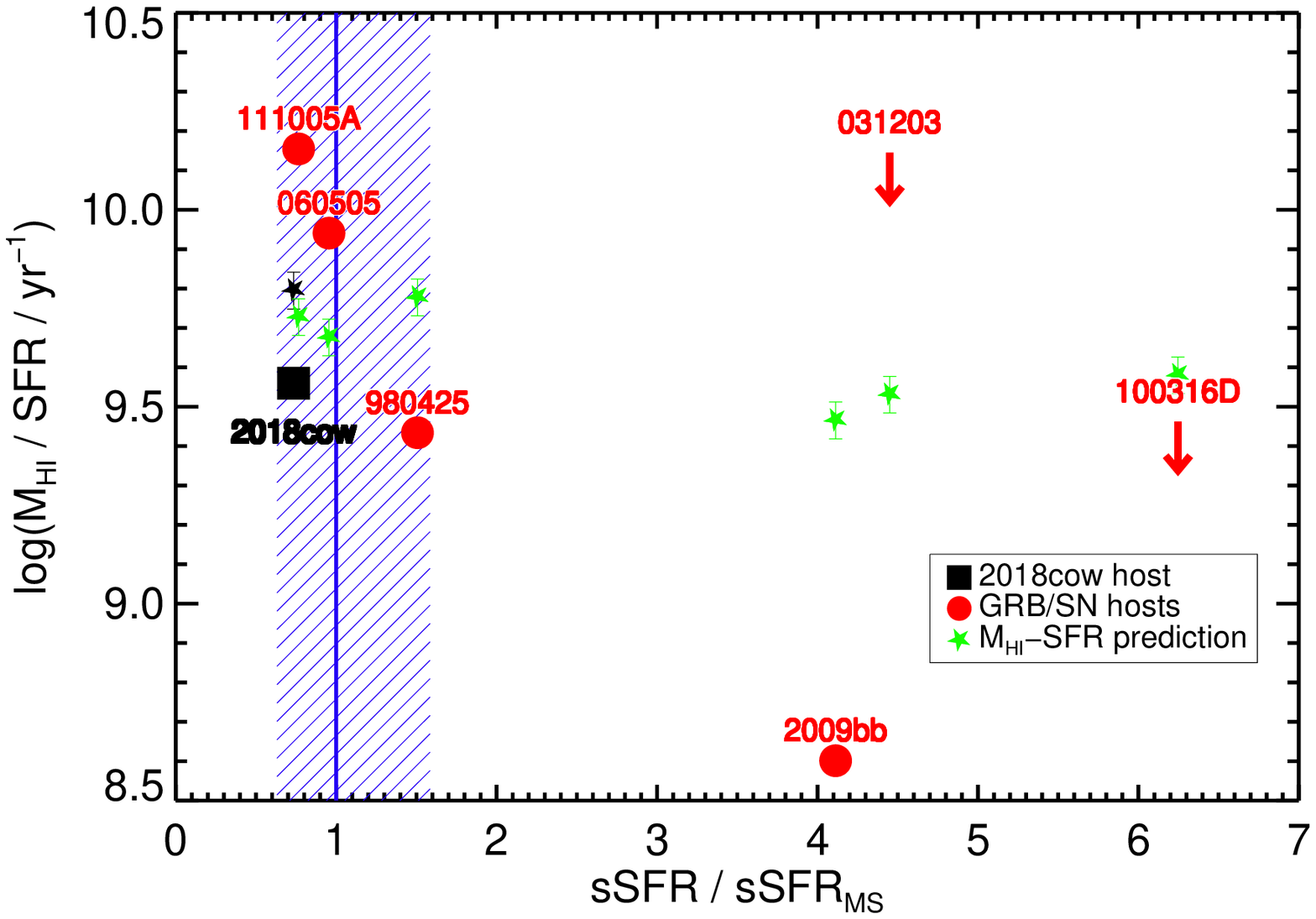}
    \caption{Atomic gas depletion timescale ($\equiv\mhi/\mbox{SFR}$) as a function of the ratio of the sSFR to the main-sequence sSFR at a given redshift and stellar mass \citep{speagle14}. {\host} and GRB/SN hosts \citep{michalowski15hi,michalowski18} are shown as a {\it black square} and {\it red circles/arrows}, respectively. The main sequence and its scatter are shown as a {\it blue vertical solid line} and the {\it hatched region}, respectively. For each galaxy, a {\it black} or {\it green star} shows the predicted gas depletion time for its SFR from the {\mhi}-SFR relation \citep{michalowski15hi}. The errors include both the uncertainty in the SFRs and in the parameters of the relation. 
    }
    \label{fig:ssfrmhi}
\end{figure*}

The {\hi} fluxes at each frequency element were determined by aperture photometry with an aperture radius of $45''$. 
The spectra are shown in Fig.~\ref{fig:hispec}. The {\hi} emission maps derived from the collapsed cubes within the dotted lines in Fig.~\ref{fig:hispec}
are shown in  Fig.~\ref{fig:all}. This range was selected to encompass the full velocity width of the line.
It was also used to obtain integrated {\hi} emission ($F_{\rm int}$ in Jy\,\kms) directly from the spectra. The line luminosity ({\lphi} in K\,\kms pc$^2$) was calculated using Eq.~3 in \citet{solomon97} and transformed to {\mhi} using Eq.~2 in \citet{devereux90}. 
The {\hi} zeroth and first moment maps (integrated emission and velocity field) are also shown on Fig.~\ref{fig:all}. 

We detected and resolved the {\hi} emission of the host of {\atcow}. The atomic gas disk is larger than the stellar disk with a centre (moment 0 `centre of mass') offset from the optical centre by 
$\sim1$--$2\arcsec$ ($\sim0.3$--$0.6\,$kpc in projection),
and $\sim5$--$6\arcsec$ or $\sim1.4$--$1.7$\,kpc from the position of {\atcow}.
Using the formula of \citet{ivison07}\footnote{$r=0.6\times\mbox{FWHM}_{\rm beam}/(\mbox{S}/\mbox{N})$, where $\mbox{FWHM}_{\rm beam}$ is the FWHM of the beam and $\mbox{S}/\mbox{N}$ is the signal-to-noise ratio.}
the positional uncertainty is $\sim0.5$--$1.5$\arcsec, so the offset to the galaxy centre is at most $2\sigma$, but to the {\atcow} position it is significant at $\sim5\sigma$.

 The {\hi} maps (Fig.~\ref{fig:all}) do not show strong evidence of recent gas inflows.
The gas distribution is much more regular than those of the hosts of GRB\,980425 \citep{arabsalmani15b}, GRB 060505 \citep{michalowski15hi}, and SN\,2009bb \citep{michalowski18}, which exhibit strong gas concentrations close to the GRB/SN positions, away from the galaxy centres. 

On the two zeroth moment maps with the highest resolutions we see the ring-like structure reported by \citet{roychowdhury19}. In Sect.~\ref{sec:discussion} we provide evidence that this structure is of internal origin.

On the outskirts of {\host} there are gas plumes in both the zeroth moment and the collapsed maps (Fig.~\ref{fig:all}), but they are of low signal-to-noise, so they cannot be interpreted as real structures with confidence. 
Moreover, they can be spiral structures. 

Moreover, from one resolution to another the `centre of mass' of the zeroth moment maps moves only by $\sim1\arcsec$ ($\sim3\arcsec$ for the $28\arcsec$ map with the worst positional uncertainty). This suggests that the distribution is symmetric.
The velocity fields (bottom row of Fig.~\ref{fig:all}) and the  double-horn profiles of the {\hi} spectra (Fig.~\ref{fig:hispec}) are consistent with a rotating disk. 

The SFR-{\mhi} relation \citep[Eq.~1 in][]{michalowski15hi}, predicts $\log(\mhi/\msun)=9.14^{+0.04}_{-0.07}$ for $\mbox{SFR}=0.22\,\msunyr$ of {\host} (the errors include both the uncertainty in the SFR and in the parameters of the relation). This is $0.24$\,dex, i.e.~$\sim3\sigma$, higher than the measured value (Table~\ref{tab:mhi}), which  is within the scatter of this relation  (0.38\,dex at $1\sigma$). Hence  {\host} has a normal atomic gas content for its SFR and is located close to most gas-poor galaxies within this relation.
The relation has been established using over 1500 galaxies, also covering the SFR range relevant here.

We present the continuum map at an observed frequency of $1.397667$\,GHz in Fig.~\ref{fig:contmap}. We detected two point sources within {\host}: {\atcow} in the south-west and a second source at the north-eastern part of the bar, just outside the bulge. The positions, fluxes, and SFRs
using the conversion of \citet{bell03} are listed in Table~\ref{tab:cont}. For the first source we show the SFR as an upper limit, as it is dominated by {\atcow}. 
This is motivated by a small offset ($0.24\arcsec$) of this source to the VLBI position of {\atcow} and the fact that a variable $1.4\,$GHz flux at this level has been reported by \citet{margutti19}.

The SFR of the second source ($0.081\,\msunyr$) is $37^{+7}_{-8}$\% of the total SFR of the galaxy as measured from the spectral energy distribution modelling \citep[$0.22\,\msunyr$;][]{perley19}. 
This source is coincident with one of the peaks of the {\hi} maps. The analogous continuum source is not  present in the other half of the bar on the other side of the bulge.
Star formation  along the bar and  differences between the two halves of the bar are common among local spirals, but regions inside bars do not dominate the total SFR \citep{regan96,sheth00,sheth02,koda06,momose10,hirota14,yajima19}. Moreover, barred spirals always exhibit significant star formation in the galaxy centre, which is not evident for {\host}.


The SFR and stellar mass of {\host} \citep{perley19} imply a specific SFR ($\mbox{sSFR}\equiv\mbox{SFR}/\mstar$) of $\sim0.15\,\mbox{Gyr}^{-1}$. At this stellar mass, the sSFR of a main-sequence galaxy is $\sim0.2\,\mbox{Gyr}^{-1}$ \citep{speagle14}. Hence, {\host} is a main-sequence galaxy at the bottom of the scatter of this relation  with no enhancement or strong suppression of star formation.

The atomic gas and star formation properties of {\host} are summarised in Fig.~\ref{fig:ssfrmhi} and compared with GRB/SN hosts with {\hi} measurements \citep{michalowski15hi,michalowski18}.
For each galaxy we also show the predicted gas depletion time from the {\mhi}-SFR relation \citep{michalowski15hi}.
GRB/SN hosts occupy two regions of this diagram: either on/below the main-sequence and abundant with atomic gas (high gas depletion timescale well above the prediction), or above the main-sequence with low gas depletion timescale, due to elevated SFR. In contrast, {\host} is below the main sequence, but it has  lower gas content than predicted from the {\mhi}-SFR relation. 
In particular it is different than the hosts of GRB\,060505 and 111005A, which have 0.3--0.5\,dex more atomic gas than predicted from their SFR. In terms of the {\mhi}/SFR ratio, the {\atcow} host is most similar to the GRB\,980425 host, which is, however, at the upper boundary of the main sequence and exhibits a strong gas concentration close to the GRB position, unlike the {\atcow} host.

\section{Discussion}
\label{sec:discussion}

The atomic gas distribution of {\host}  does not show strong unusual features (especially not at the {\atcow} position), in contrast to the off-centre gas concentrations and irregular velocity fields of the host galaxies of
GRBs or relativistic SNe \citep{arabsalmani15b,michalowski14,michalowski15hi,michalowski16,michalowski18}. Moreover, there is no enhancement of the SFR, which could be a signature of a gas inflow. 
The environment of {\atcow} therefore suggests that its progenitor may not have been a massive star
\citep{prentice18,margutti19, riverasandoval18,fox19}. However, the GRB/SN host sample with atomic gas measurements is small, so we cannot rule out this hypothesis.

The asymmetry in the distribution of atomic gas in the case of the host of the relativistic SN\,2009bb may be a result of interaction \citep{michalowski18}, as has also been observed for other galaxies \citep{sancisi08,rasmussen06}. Hence to investigate this further we analysed the large-scale environment of {\host} using the NASA/IPAC Extragalactic Database (NED) of 
 {\host}. It seems fairly isolated, with no other galaxies within 500\,kpc projected distance and 1000\,{\kms} velocity. The nearest galaxy is UGC\,10322, more than 500\,kpc away in projected distance. Hence, in the current catalogues there is no galaxy which is close enough to significantly influence the properties of {\host}. We found that {\host} is $\sim700\,$kpc to the west of a possible galaxy group extending several hundred kpc across and containing six galaxies.
Similarly to the host of SN\,2009bb \citep{michalowski18}, this could mean that there is  a supply of intergalactic gas available for
inflow onto {\host}, but we did not find any evidence of such process.

On the other hand, all other proposed explosion mechanisms of {\atcow}, apart from massive-star core-collapse, should not result in a connection between its progenitor and gas concentration or inflow: 
an exploding low-mass hydrogen-rich star \citep{margutti19},
a TDE \citep{liu18,kuin19,perley19},
and a merger of white dwarfs or a neutron star and a giant star \citep{lyutikov19,soker19}.
Hence, the normal atomic gas distribution of {\host} is consistent with these mechanisms.

After the submission of this paper, the results of  \citet{roychowdhury19} on the atomic gas distribution in the host galaxy of {\atcow} were published. They found a ring of gas, also visible in our combined dataset (Fig.~\ref{fig:all}).

As claimed by \citet{roychowdhury19}, such a gas ring could be the result of a minor merger. However, most  rings in galaxies has been shown to be the result of resonances caused by the presence of a bar (gravitational torques; see the review by \citealt{buta96}) and other internal mechanisms, like viscous torques \citep{icke79,buta86,lesch90,combes85,armillotta19}. Similarly, \citet{diazgarcia19} found an increasing fraction of ringed galaxies with increasing  bar Fourier density
amplitude (also for galaxies with stellar masses similar to that of {\host}).

Indeed {\host} exhibits a strong bar, which can be the cause of the appearance of the gas ring. Moreover, the {\hi} velocity fields presented here and by \citet{roychowdhury19} do follow a rotation pattern, and do not show any sign of disturbances, given the errors in the measurements.
Finally, almost all spiral galaxies (including those with similar masses to {\host}) exhibit central depressions of atomic gas (likely due to conversion to the molecular phase) or enhancement at the location of the spiral arms \citep{leroy08,bigiel12,martinsson16}. This feature, combined with low sensitivity \citep[as in the highest-resolution map of][]{roychowdhury19} would give rise to a ring-like structure in the data, which would have a purely internal origin.
Hence, the presence of the gas ring in {\host} without any sign of disturbance is not  strong evidence of a recent merger.

\section{Conclusions}
\label{sec:conclusion}

We observed the {\hi} atomic hydrogen line emission of the {\atcow} host galaxy with the Giant Metrewave Radio Telescope.
There is no unusual atomic gas concentration near the position of {\atcow}.
The gas distribution is much more regular than those of the hosts of GRBs and SNe. The {\atcow} host has an atomic gas mass lower by 0.24\,dex than the prediction from its SFR and is at the lower edge of the galaxy main sequence. 
In the continuum we detected the emission of {\atcow} and of a star-forming region in the north-eastern part of the bar (away from {\atcow}). This region hosts a third of the galaxy star formation rate (SFR).

The absence of atomic gas concentration close to {\atcow}, along with a normal SFR and regular {\hi} velocity field  sets CGCG137-068 apart from GRB/SN hosts studied in {\hi}. 
The environment of {\atcow} therefore suggests that its progenitor may not have been a massive star.
Our findings are consistent with  an origin of the transient that does not require a connection between its progenitor and gas concentration or inflow: 
an exploding low-mass star, a tidal disruption event, 
or a merger of white dwarfs or of a neutron star and a giant star.
We interpret the recently reported atomic gas ring in CGCG\,137-068 as a result of internal processes connected with gravitational resonances caused by the bar.

\begin{acknowledgements}

We thank Joanna Baradziej for help in improving this paper.

M.J.M.~acknowledges the support of 
the National Science Centre, Poland, through the POLONEZ grant 2015/19/P/ST9/04010 and SONATA BIS grant 2018/30/E/ST9/00208;
this project has received funding from the European Union's Horizon 2020 research and innovation programme under the Marie Sk{\l}odowska-Curie grant agreement No. 665778.
P.K. is supported by the BMBF project 05A17PC2 for D-MeerKAT
J.H. was supported by a VILLUM FONDEN Investigator grant (project number 16599).
D.A.K. acknowledges support from the Juan de la Cierva Incorporaci\'on fellowship IJCI-2015-26153.
A.d.U.P.~and C.C.T.~acknowledge support from Ram\'on y Cajal fellowships (RyC-2012-09975 and RyC-2012-09984). D.A.K., A.d.U.P., and C.C.T.~acknowledge support from the Spanish research project AYA2017-89384-P.
L.K.H.~acknowledges funding from the INAF PRIN-SKA program 1.05.01.88.04. The Cosmic Dawn Center is funded by the DNRF.
R.L. acknowledges support from the grant EMR/2016/007127 from the Dept. of Science and Technology, India.
This project has received funding from the European Research Council (ERC) under the European Union’s Horizon 2020 research and innovation programme (grant agreement no. 679627; project name FORNAX; PI Paolo Serra).

We thank the staff of the GMRT who have made these observations possible. GMRT is run by the National Centre for Radio Astrophysics  of the Tata Institute of Fundamental Research. 
Based on observations made with the Gran Telescopio Canarias (GTC), in the Roque de los Muchachos Observatory.
We acknowledge the usage of the HyperLeda database (\url{http://leda.univ-lyon1.fr}).
This research has made use of 
the NASA/IPAC Extragalactic Database (NED), which is operated by the Jet Propulsion Laboratory, California Institute of Technology, under contract with the National Aeronautics and Space Administration;
SAOImage DS9, developed by Smithsonian Astrophysical Observatory \citep{ds9};
and  NASA's Astrophysics Data System Bibliographic Services. 

\end{acknowledgements}




\end{document}